\documentclass [12pt]{article}

\usepackage{latexsym}
\usepackage{epsfig}

\newcommand {\Z}{{\bf Z}}
\newcommand {\C}{{\bf C}}
\newcommand {\N}{{\bf N}}
\newcommand {\R}{{\bf R}}

\newcommand {\cL}{\Lambda^C}
\newcommand {\bL}{\Lambda_2^C}

\newcommand {\e}{\epsilon}
\newcommand {\vth}{\vartheta}
\newcommand {\tvth}{\tilde{\vartheta}}
\newcommand {\beq}{\begin {equation}}
\newcommand {\eeq}{\end {equation}}

\newcommand {\id}{\mbox{id}}

\newcommand {\ba}{\begin {eqnarray}}
\newcommand {\ea}{\end {eqnarray}}

\newcommand {\ad}{{A_\delta}}

\newcommand {\adz}{{A_\delta^{\Lambda_2}}}

\newcommand {\qed}{\hfill $\Box$ \vspace{0.3cm}}
\newcommand {\la}{\label}
\newcommand {\nn}{\nonumber}
\newcommand {\dist}{{\mbox{\tiny dist} (\Lambda_1 ,\Lambda_2 )}}
\newcommand {\bdist}{{\mbox{dist} (\Lambda_1 ,\Lambda_2 )}}

\title{Transfer Operators for Coupled Analytic Maps}
\author{Torsten Fischer\thanks{supported by the EC via 
TMR-Fellowship ERBFMBICT-961157} and Hans Henrik Rugh\\
Mathematics Institute\\University of Warwick\\Coventry CV4 7AL U.K.}

\begin {document}

\newtheorem{theorem}{Theorem}
\newtheorem{proposition}{Proposition}
\newtheorem{lemma}{Lemma}

\maketitle



\begin{abstract}
We consider analytic coupled map lattices over $\Z^d$ with exponentially
decaying interaction. 
We introduce Banach spaces for the infinite-dimensional system that
include measures with analytic, exponentially bounded finite-dimensional 
marginals.
Using residue calculus and `cluster expansion'-like 
techniques we define transfer operators on 
these Banach spaces.
For these we get a unique probability measure that exhibits exponential
decay of correlations.
\end{abstract}


\setcounter{section}{-1}
\section{Introduction}

Coupled map lattices were introduced by K. Kaneko (cf.\ \cite{kaneko} for
a review)
as systems that are weak mixing wrt.\ spatio-temporal shifts.
L.A. Bunimovich and Ya.G. Sinai proved in \cite{bunimovichsiani}
(cf.\ also the remarks on that in \cite{bricmontkupiainen2}) the existence
of an invariant measure and its exponential decay of correlations for a
one-dimensional lattice of weakly coupled maps by constructing a 
Markov partition and relating the system to a two-dimensional spin system.

J. Bricmont and A. Kupiainen extend this result in
\cite{bricmontkupiainen1} and \cite{bricmontkupiainen2,bricmontkupiainen3}
to coupled circle maps over the $\Z^d$-lattice with analytic and
H\"{o}lder-continuous weak interaction, respectively. They use a `polymer' or
`cluster'-expansion for the Perron-Frobenius operator for the 
finite-dimensional subsystems over $\Lambda \subset \Z^d$ and write
the $n$th iterate  of this operator applied to the constant function $1$
in terms of potentials for a $d+1$-dimensional spin system.
Taking the limit as $n \to \infty$ and $\Lambda \to \Z^d$ they get
existence and uniqueness (among measures with certain properties)
of the invariant 
probability measure and exponential decay of correlations.

V. Baladi, M. Degli Esposti, S. Isola, E. J\"{a}rvenp\"{a}\"{a}
and A. Kupiainen
define in \cite{baladi}, 
for infinite-dimensional systems
over the $\Z^d$ lattice, transfer operators on a Frechet space, and, 
for $d=1$, on a Banach space; they study the spectral properties of 
these operators, viewing the coupled operator as a perturbation 
of the uncoupled one in the Banach case.

In \cite{kellerkuenzle} G. Keller and M. K\"{u}nzle consider periodic
or infinite one-dimensional lattices of weakly coupled maps of the unit
interval. In particular they define transfer operators on the space $BV$
of measures whose finite-dimensional marginals are of bounded variation
and prove the existence of an invariant probability measure.
For the infinite-dimensional system they further show that for a small
perturbation of the uncoupled map any invariant measure in $BV$ 
is close (in a specified sense) to the one they found.

Coupled map lattices with multi-dimensional local systems of hyperbolic type
have been studied by Ya.B. Pesin and Ya.G. Sinai \cite{pesinsinai},
M. Jiang \cite{jiang1,jiang2}, M. Jiang and A. Mazel \cite{jiangmazel},
M. Jiang and Ya.B. Pesin \cite{jiangpesin} and
D.L. Volevich \cite{volevich1,volevich2}.

More detailed surveys on coupled map lattices can be found in
\cite{bunimovich}, \cite{jiangpesin} and \cite{bricmontkupiainen2}.

\vspace{0.5cm}
In the above papers (except \cite{baladi}, \cite{kellerkuenzle}) 
the analysis has been done only
for Banach spaces defined for finite subsets $\Lambda$ of the lattice,
and the (weak) limit of the invariant measure for $\Lambda \to \Z^d$
was taken afterwards.

Here we present a new point of view in which a natural
Banach space and transfer operators are defined
for the infinite lattice of weakly coupled analytic maps 
(Section \ref{sectiongeneralsetting}).
The space contains consistent families of analytic marginals over
finite subsets of $\Z^d$.
We take a weighted sup-norm
so that the sup-norms of the marginals for the sub-systems over finitely many
(say $N$) lattice points is bounded exponentially in $N$ 
(Section \ref{sectionmainresults}).
We identify an ample subset of this space with a set of \emph{rca} measures
(Section \ref{sectionfurtherremarksontheinfinitedimensionalsystem})
that contains the unique invariant probability density
(Section \ref{sectionmainresults}).

We derive exponential decay of correlations for this measure
from (the proof of) the spectral properties of our transfer operators.
(Sections \ref{sectionmainresults}, \ref{sectiondecayofcorrelations}).

Our approach provides a natural setting for an analysis of the full
$\Z^d$ Perron-Frobenius operator in terms of cluster expansions
over finite subsets of the lattice.
Using resi\-due calculus we introduce an integral representation for the
Perron-Frobenius operator for finite-dimensional sub-systems 
(Section \ref{sectionfinitedimensionalsystems}) which yields a uniform 
control over the perturbation
and also gives rise 
to an easy approach to stochastic perturbation (cf.\ \cite{maesmoffaert})
which however we do not consider here.

Our `cluster expansion' combinatorics
(Section \ref{expansionoftheperronfrobeniusoperator}) uses ideas from
\cite{maesmoffaert} (cf.\ also \cite{bricmontkupiainen1}).
Apart from the analysis of the one-dimensional operator, which is fairly
standard and for which we refer
to e.g.\ \cite{bricmontkupiainen1}, the paper should be self-contained.


\section{General Setting}
\la{sectiongeneralsetting}

We consider coupled map lattices in the following setting:
The state space is $M = (S^1)^{\Z^d}$ where 
$S^1 =\{ z\in \C \mid \left| z \right| =1\} $
is the unit circle in the complex plane and $d\in N$.

The map $ S : M \rightarrow M$ is the composition $ S=F\circ T^{\e}$
of a coupling map $T^{\e }$ depending on a (small) non-negative parameter
$\e $ and another parameter for the decay of interaction (cf.\ (\ref{formgpk}))
with an (uncoupled) map $F$ that acts on each  component of M separately. 
We make the following assumptions:

\vspace{0.3cm}
{\bf I } 
$F ({\bf z})=(f_p(z_p))_{p\in {\Z^d}}$ 
where $f_p: S^1 \rightarrow S^1$ are real analytic and expanding
(i.e.\ $f_p' \geq {\lambda }_0 > 1$) maps that extend for some $\delta_1$
holomorphically
to the interior of an annulus $A_{\delta_1}\stackrel{\mathrm{def}}{=} 
\{z\in \C \mid -{\delta}_1 \leq \ln \left| z \right| \leq{\delta}_1\}$ and 
the family of Perron-Frobenius operators $\mathcal{L}_{f_p}$ 
for the indiviual
systems satisfies uniformly a condition specified in Section
(\ref{subsectionunperturbedoperator}).

\vspace{0.3cm}
We write $T^{\e} : M \rightarrow M$ as $T^{\e} ({\bf z})=
(T^{\e}_p ({\bf z}))_{p\in {\Z^d}}$ and
$T^{\e}_p ({\bf z})= z_p \exp [2\pi\imath\e g_p({\bf z})]$ with
$g_p({\bf z}) = \sum_{k=1}^{\infty} g_{p,k}({\bf z})$.
The functions
$g_{p,k}$ is real valued on $(S^1)^{\Z^d}$ and
depends only on those $z_q$ with $\| p-q \|
\leq k$ (neighbours of distance at most k) where 
$\| p\| \stackrel{\mathrm{def}}{=} \sum_{l=1}^{d} | p_l |$. 

We write  $B_k(p)=\{ q\in \Z^d \mid \| p-q\| \leq k\}$
and also denote by $g_{p,k}$  the function from the finite-dimensional
torus $(S^1)^{B_k(p)}$ to $\R$.

We assume the following for the functions $g_{p,k}$:

\vspace{0.3cm}
{\bf II} 
For all
$p \in \Z^d$ and $k \geq 1$ the maps $g_{p,k}$ extend to a 
holomorphic map $g_{p,k} : A_{{\delta}_1}^{B_k(p)} \rightarrow \C$
and
\beq
{\parallel g_{p,k} \parallel}_{A_{\delta_1}^{B_k(p)}} \leq 
c_1 \exp \left( -c_2 k^d \right)
\la{formgpk}
\eeq 
with $c_1>0$ and
$c_2$ bigger than a certain constant specified in (\ref{formconditionc2}).

\vspace{0.3cm}
The parameter $c_1$ is actually redundant as it
is multiplied by $\e$ in the definition of $T_p^{\e}$. We also have
$\exp (-c_2 k^d)\leq
\exp (-\xi ) \exp (-\tilde{c}_2 k^d)$
for  $\tilde{c}_2 = c_2 - \xi ,\, \xi >0$, i.e.\ for any $\e$
we can make the interaction 
small only by taking $c_2$ large. But once we have chosen $c_2$ large
enough to guarantee the convergence of the infinite sums in our analysis
we can consider perturbations of the uncoupled map 
depending on the parameter $\e$ only.

With the metric

\beq
d_{\gamma}({\bf x},{\bf y})\stackrel{\mathrm{def}}{=} 
\sup_{p\in \Z^d} \gamma^{\| p\|} \| x_p-y_p\|
\eeq

for $0<\gamma <1 \quad$
$(M,d_{\gamma})$ is a compact metric space.
Its topology is the product topology on $\left( S^1\right)^{\Z^d}$.
The Borel $\sigma$-algebra $\mathcal{B}$ on $M$ is the same as the 
product $\sigma$-algebra.
$F$ and $T^\e$ are continuous and measurable.
Let $\mathcal{C}(M)$ denote the space of real-valued continuous functions on
$(M,d_{\gamma})$ with the sup-norm and $\mu$ the 
Lebesgue (product) measure on $M$.

For $\Lambda_1 \subseteq \Lambda_2 \subseteq \Z^d$, with
$\Lambda_1$ finite and an integrable function $g$ on $M$ depending only on
the $\Lambda_2$-coordinates, we define the projection

\beq
(\pi_{\Lambda_1} g) ({\bf z}_{\Lambda_1} )
\stackrel{\mathrm{def}}{=} 
\int_{(S^1)^{\Lambda_2 \setminus
\Lambda_1}} d\mu^{\Lambda_2 \setminus \Lambda_1}
({\bf z}_{\Lambda_2 \setminus \Lambda_1}) 
g({\bf z}_{\Lambda_1} \vee {\bf z}_{\Lambda_2 \setminus \Lambda_1})
\eeq


\section{Main Results}
\la{sectionmainresults}

For finite $\Lambda \subset \Z^d$ let
$H(A_\delta^{\Lambda})$ be the space of continuous functions
on the closed polyannulus $A_\delta^{\Lambda}$ that are holomorphic
on its interior 
and write
$\|\cdot\|_\Lambda$ for the sup-norm on $H(A_\delta^\Lambda)$.
Let $\mathcal{F}$ be the set of all finite subsets (including
$\emptyset$)  of $\Z^d$. 
We denote
by $\mathcal{H}$ the set of all consistent families
$\phi =(\phi_{\Lambda} )_{{\Lambda}\in\mathcal{F}}$ 
of functions
$\phi_{\Lambda} \in H(A_\delta^{\Lambda} )$.
Consistency means $\pi_{\Lambda_1} \phi_{\Lambda_2}=\phi_{\Lambda_1}$ for
$\Lambda_1 \subseteq \Lambda_2 \in \mathcal{F}$. 
We write $\mu (\phi ) \stackrel{\mathrm{def}}{=}\phi_\emptyset$.

We want to define a norm on a (sufficiently large)
subspace of $\mathcal{H}$ that should at least contain
`product densities' like $h=(h_\Lambda)_{\Lambda\in\mathcal{F}}$ 
with $h_\Lambda ({\bf z})= \prod_{p\in\Lambda} h_p(z_p)$, where
$h_p \in H(A_\delta^{\{ p\} } )$ is the invariant probability density
for the single system over $\{ p\}$ (cf.\ Section
\ref{subsectionunperturbedoperator}).
As $\| h_p\|_{\{ p\} } \leq c_h$ uniformly in $p$,
the sup-norm
$\| h_{\Lambda_1} \|_{\Lambda_1}$ does not grow faster than
exponentially in $|\Lambda_1 |$.
Therefore we take a weighted sup-norm.
For $0<\vth <1$ 
we define

\beq
\|\phi\|_{\vth} = 
\sup_{\Lambda \in \mathcal{F}}
\vth^{|\Lambda |} \|\phi_{\Lambda}\|_{\Lambda}
\eeq

and set $\mathcal{H}_{\vth} \stackrel{\mathrm{def}}{=}
\left\{ \phi\in \mathcal{H} \mid 
\|\phi\|_{\vth }< \infty \right\}$.
Then $\left(\mathcal{H}_{\vth },\|\cdot\|_{\vth }
\right)$ is a 
Banach space.
Analogously we define for $\Lambda \in \mathcal{F}$
the weighted norm on spaces 
$\mathcal{H}_{\Lambda , \vth}$ 
of consistent sub-families $\left( \phi_{\Lambda_1}
\right)_{ 
\Lambda_1 \subseteq \Lambda}$: 

\beq
\|\phi\|_{\Lambda , \vth} \stackrel{\mathrm{def}}{=} 
\sup_{\Lambda_1\subseteq \Lambda}
\vth^{|\Lambda_1 |} \|\phi_{\Lambda_1} \|_{\Lambda_1}
\eeq
 
We get the same 
(topological) vector space as $\left( H(A_\delta^{\Lambda} ),
\|\cdot\|_{\Lambda}\right)$, but the constants for
the estimates of the norms are unbounded as $| \Lambda |$ increases.

For given $\Lambda_1 \subseteq \Lambda_2 
\in \mathcal{F}$ and $N\in \N$ we have a map

\beq
\pi_{\Lambda_1} \circ \mathcal{L}^N_{F^{\Lambda_2} 
\circ T^{\Lambda_2 ,\e}} \circ \pi_{\Lambda_2} :
\left(\mathcal{H}_{\vth} ,\|\cdot\|_{\vth}
\right) \rightarrow
\left(\mathcal{H}_{\Lambda_1 ,\vth},\|\cdot\|_{\Lambda_1 ,
\vth}\right)
\la{formpfoinf2}
\eeq

where $\mathcal{L}^N_{F^{\Lambda_2} \circ T^{\Lambda_2 ,\e}}$
is the Perron-Frobenius operator for the finite-dimensional 
system over $\Lambda_2$ (cf.\ Section \ref{sectionfinitedimensionalsystems})
with fixed boundary conditions (not included in the notation).
The following definition of transfer operators for the 
infinite system does not depend on the choice of the boundary conditions.


\begin{theorem}
For $\vth$, $\e$ sufficiently small, $c_2$, $N$ sufficiently big 
and any $\Lambda_1 \in \mathcal{F}$: 
\la{theoremexistencepfo}
\end{theorem}

{\itshape
\begin{enumerate}
\item
The limit
\beq
\pi_{\Lambda_1} \circ \mathcal{L}^N_{F \circ T^\e}
\stackrel{\mathrm{def}}{=} 
\lim_{\Lambda_2 \rightarrow \Z^d}
\pi_{\Lambda_1} \circ \mathcal{L}^N_{F^{\Lambda_2} 
\circ T^{\Lambda_2 ,\e}} \circ \pi_{\Lambda_2}
\la{formpi1pfo}
\eeq

$\in 
L\left(\left(\mathcal{H}_{\vth} ,
\|\cdot\|_{\vth}\right),
\left( \mathcal{H}_{\Lambda_1 ,\vth} ,
\|\cdot\|_{\Lambda_1 ,\vth}\right)\right)$
exists and the family of these operators is uniformly (in
$\Lambda_1$ and also in $N$) bounded. 
This defines operators $\mathcal{L}^N_{F \circ T^\e}$
on $\left(\mathcal{H}_{\vth} ,\|\cdot\|_{
\vth}\right)$ by
$\left(\mathcal{L}^N_{F \circ T^\e} \phi \right)_{\Lambda_1}
\stackrel{\mathrm{def}}{=}
\pi_{\Lambda_1} \circ \mathcal{L}^N_{F \circ T^\e} \phi $

For any $n\in \N$ we have
$\mathcal{L}^n_{F \circ T^\e}:
\mathcal{H}_\vth \rightarrow \mathcal{H}_{\vth_n}$ 
with suitably chosen
$0< \vth_1 \leq \cdots \leq \vth_{N_0}=\vth_{N_0+1}=\cdots =\vth$.

In the case of finite-range interaction we can define 
a linear map $\mathcal{L}_{F \circ T^\e}$ on
$\mathcal{H}$ in the
same way .

\item
There is a unique invariant probability measure 
$\nu =(\nu_{\Lambda_1} )_{\Lambda_1 \in\mathcal{F}}\in
\mathcal{H}_{\vth}$.

In 
$L\left(
\mathcal{H}_{\vth},\|\cdot\|_{\vth}\right)$
the sequence
$\left( \mathcal{L}^N_{F \circ T^\e}
\right)_{N \geq N_0}$
converges exponentially\\ fast:

\beq
\left\| \mathcal{L}^N_{F \circ T^\e} - \mu (\cdot ) \nu
\right\|_{L
\left(\left(\mathcal{H}_{\vth} ,\|\cdot\|_{\vth}
\right)\right)} 
\leq c_3 \tilde{\eta}^N
\la{formestimatepfoNminusmu}
\eeq
for some $c_3 >0$ and $0 < \tilde{\eta} <1$.
\end{enumerate}}

For the invariant measure $\nu$ we have exponential decay of correlations
for spatio-temporal shifts on the system:

Let $(e_1,\ldots ,e_d)$ be a linearly-independent system of
unit vectors in $\Z^d$.
We define translations 
$\tau_{e_i} (p) \stackrel{\mathrm{def}}{=} p+ e_i$ for $p\in \Z^d$
and $(\tau_{e_i}(z))_p \stackrel{\mathrm{def}}{=} z_{\tau_{e_i} (p)}$
for $z \in M$.

In the following theorem we denote by $\tau$ (acting on $M$ 
from the right) compositions
$ \tau = \tau_{1}\circ\ldots\circ\tau_{m(\tau )}$
and by $\sigma$ a composition of  spatio-temporal shifts (on $M$):
$\sigma =\sigma_1 \circ\ldots\circ\sigma_{m(\sigma ) + m(\sigma )}$ with
$\sigma_i \in \{ S,\tau_{e_1},\ldots ,\tau_{e_d}\}$. We denote by
$n(\sigma)$
the number of factors $S$ and by $m(\sigma )$ the number of
spatial translations in this product.
For a translation-invariant system, i.e.\ $f_p=f$ 
and $g_p({\bf z})=g_{\tau_{e_i}^{-1} (p)}(\tau_{e_i} (z))$ for all 
$p\in \Z^d$,
the time-shift $S$ commutes with the translations. 


\begin{theorem}
For $\vth$,$\e$ and $c_2$ as in Theorem \ref{theoremexistencepfo}
there is a
$\kappa\in (0,1)$ such that for all
nonempty $\Lambda_1 , \Lambda_2 \in \mathcal{F}$ with
$\Lambda_1 \cap \Lambda_2 =\emptyset$ the following holds:
\la{theoremdecay2}
\end{theorem}

{\itshape
\begin{enumerate}

\item
If $g\in \mathcal{C}((S^1)^{\Lambda_1})$ and
$f\in \mathcal{C}((S^1)^{\Lambda_2})$ then

$\left| \int_M \nu d\mu \, g f - 
\left( \int_M \nu d\mu \, g \right)
\left( \int_M \nu d\mu \, f \right) \right|
\leq
c_4 \vth^{-|\Lambda_1| - |\Lambda_2|} \| g\|_\infty
\| f\|_\infty \kappa^{\dist}$


\item
If $g\in \mathcal{C}((S^1)^{\Lambda_1})$
and $f \in \mathcal{H} \cap \mathcal{C}((S^1)^{\Lambda_2})$
then

\ba
\lefteqn{\left| \int_M \nu d\mu \, g\circ\tau\circ S^n f - 
\left( \int_M \nu d\mu \, g\circ\tau \right)
\left( \int_M \nu d\mu \, f \right) \right|} \\
& \leq &
c(\Lambda_1 ,\Lambda_2 ,\kappa) c_5^{|\Lambda_1|+|\Lambda_2|} 
\| g\|_\infty
\| f\|_{\Lambda_2} \kappa^{m(\tau )} \tilde{\eta}^n
\nn
\ea

with suitable $c_5$ and $\tilde{\eta}$ as in Theorem
\ref{theoremexistencepfo}.


\item
If the system is translation-invariant and $g,f$ are as in (2.\ ), then

\ba
\lefteqn{\left| \int_M \nu d\mu \, g\circ\sigma f - 
\left( \int_M \nu d\mu \, g \right)
\left( \int_M \nu d\mu \, f \right) \right|}\\
& \leq &
c(\Lambda_1 ,\Lambda_2 ,\kappa) 
c_5^{|\Lambda_1| + |\Lambda_2|} \| g\|_\infty
\| f\|_{\Lambda_2}  
\kappa^{m(\sigma )} \tilde{\eta}^{n(\sigma )} 
\nn
\ea


\item
If $g,f\in \mathcal{C}(M)$ then

\beq
\lim_{\max \{ m(\tau ),n\} \to\infty}
\left| \int_M \nu d\mu \, g\circ\tau\circ S^n f -
\left( \int_M \nu d\mu \, g\circ\tau \right)
\left( \int_M \nu d\mu \, f \right)\right| =0  .
\eeq


\item
If the system is translation-invariant and $g,f\in \mathcal{C}(M)$ then

\beq
\lim_{\max \{ m(\sigma ),n(\sigma )\} \to\infty}
\int_M \nu d\mu \, g\circ\sigma \, f 
=
\left( \int_M \nu d\mu \, g \right)
\left( \int_M \nu d\mu \, f \right)
\eeq	

\end{enumerate}}

{\bf Remarks: }1)Statement (5.\ ) means that for a translation-invariant 
system $\nu$ is mixing wrt.\ spatio-temporal shifts. According to (3.\ ),
the decay of correlations for observables $g$ and $h$ as specified in (2.)
is exponentially fast.

2) We could choose the rate of decay $\kappa$ first and
then the other parameters.

3)The integration wrt.\ `$\nu d \mu$' will be defined in
Section \ref{sectionfurtherremarksontheinfinitedimensionalsystem}.

4) $c(\Lambda_1 ,\Lambda_2 ,\kappa)$ in (2.) and (3.) is a constant 
depending only on
$\mbox{dist} (\Lambda_1 ,\Lambda_2 )$ and $\kappa$.


\section{Finite-dimensional Systems}
\la{sectionfinitedimensionalsystems}
 
We first consider `finite-dimensional versions' of the maps 
$F,T^{\e}$ etc.
For a finite subset $ \Lambda \in \Z^d$  and some fixed configuration
$ {\bf z}_{\cL}=  (z_p)_{p\in {\cL}} \subset
(S^1)^{\cL}$ 
on $\cL \stackrel{\mathrm{def}}{=} \Z^d \setminus \Lambda$
we define
$T^{\Lambda , \e}: A_{\delta}^{\Lambda} \rightarrow \C^{\Lambda}$
by
$(T^{\Lambda , \e}({\bf z}_{\Lambda}))_p \stackrel{\mathrm{def}}{=}
z_p \exp (2\pi\imath\e g_p ({\bf z}_{\Lambda}\vee {\bf z}_{\cL}))$,
where ${\bf z}_{\Lambda}\vee {\bf z}_{\cL}\in M$ agrees with
${\bf z}_{\Lambda}$ on its $\Lambda$-sites and with 
${\bf z}_{\cL}$
on its $\cL$-sites.

We do not specify ${\bf z}_{\cL}$ in the notation of
$T^{\Lambda , \e}$. The restriction of F to 
$A^{\Lambda}_{\delta}$ is 
denoted by $F^{\Lambda}$.

With the following two propositions we ensure that for sufficiently small
$\delta$ and $\e$ (depending on $\delta$ but not on $\Lambda$ or 
${\bf z}_{\cL}$),
$F^{\Lambda}\circ T^{\Lambda ,\e}$ maps $A_{\delta}^{\Lambda}$
to a bigger polyannulus (cf.\ \cite{bricmontkupiainen1}).

For $\Lambda\subset \Z^d$ we have the metric $d_{\Lambda}$ on 
$(S^1)^{\Lambda}$ defined by
\beq 
d_{\Lambda} ({\bf z}, {\bf w}) \stackrel{\mathrm{def}}{=} 
\sup\{\left| z_p -w_p \right| \mid p\in \Lambda\}
\la{definitionlambdametric}
\eeq

 
\begin{proposition}

For all $c_7 \in (0,1)$, sufficiently small $\delta$ (depending on $c_7$)
and $\e$ (depending on $c_7$ and $\delta$), and arbitrary
$\Lambda \in \mathcal{F}$, \,
$T^{\Lambda , \e}$ maps $A_{\delta}^{\Lambda}$ biholomorphically onto its 
image and $T^{\Lambda , \e} \left(A_{\delta}^{\Lambda}
\right) \supset A_{c_7 \delta}^{\Lambda}$, i.e.\ the image contains a 
sufficiently thick polyannulus. Also
$T^{\Lambda , \e} \left(\partial A_{\delta}^{\Lambda}
\right) \cap A_{c_7 \delta}^{\Lambda} =\emptyset$,
i.e. the image of the boundary (the same as the boundary of the image)
does not intersect the closed smaller polyannulus. 
\la{propcd}
\end{proposition}


\begin{proposition}
Let the expanding maps
$f_p: S^1 \rightarrow  S^1$ satisfy condition ${\bf I}$ for some
$\delta_1$ and an expansion constant $\lambda_{0}$
and let $1 < \lambda < \lambda_0$.
Then for all sufficiently small $\delta$ ($0<\delta <\delta_0$)
and all finite $\Lambda\subset \Z^d$
the map
$F^{\Lambda} : A_{\delta}^{\Lambda} \rightarrow \C^\Lambda$ is locally
biholomorphic, $ A_{\lambda\delta}^{\Lambda}
\subset F^{\Lambda}\left( A_{\delta}^{\Lambda}\right)$,
i.e.\ the image contains a
thicker polyannulus and furthermore all 
${\bf z}\in A_{\lambda\delta}^{\Lambda}$ have the same number of preimages.
We also have
$ A_{\lambda\delta}^{\Lambda}
\cap F^{\Lambda}\left( \partial A_{\delta}^{\Lambda}\right) = \emptyset$.
\la{propld}
\end{proposition}

Combining Propositions \ref{propcd} and \ref{propld} we have for 
fixed $c_7$ and 
(small) $\delta$
\beq
F^{\Lambda}\circ T^{\Lambda ,\e}\left( A^{\Lambda}_{\delta}\right)
\supset A^{\Lambda}_{c_7 \lambda\delta} 
\eeq
and
\beq
F^{\Lambda}\circ T^{\Lambda ,\e}\left( \partial A^{\Lambda}_{\delta}\right)
\cap A^{\Lambda}_{c_7 \lambda\delta} = \emptyset 
\eeq

In particular, if we choose $c_7>\frac{1}{\lambda}$ there is a disc
of radius $(c_7\lambda -1)\delta >0$ 
around each point in $A^{\Lambda}_{\delta}$
that is entirely contained in 
$F^{\Lambda}\circ T^{\Lambda ,\e}\left( A^{\Lambda}_{\delta}\right)$.
We will need this for Cauchy estimates. From now on we keep
$\delta$ fixed.

In the next proposition we establish a special representation
of the Perron-Frobenius operator for our finite system with
$(S^1)^N=\left( S^1\right)^{\Lambda}$, 
$S^\e=F^{\Lambda}\circ T^{\Lambda ,\e}$,
$\psi$ continuous (the proposition holds also 
for $\psi\in L^\infty (M)$)
and $\phi$ continuous on the closed
polyannulus $A_{\delta_1}^{\Lambda}$ and analytic in its interior.

First we give the definition of the Perron-Frobenius operator 
(cf.\ for example \cite{lasotamackey}).


\vspace{0.5cm}
{\bf Definition }
Let $\mu$ be a measure on a metric space $M$ (with the Borel 
$\sigma$-algebra).
The Perron-Frobenius operator $\mathcal{L}_S$ for a nonsingular 
measurable map $S: M \rightarrow M$
is defined via the equation

\beq
\int_{M} d\mu \, \psi\circ S \: \phi = 
              \int_{M} d\mu \, \psi \, \mathcal{L}_S \phi
\la{defpfo}
\eeq

that, for given $\phi \in {L^1} (M)$, must hold for all
$\psi\in L^{\infty}(M)$. 
The existence and uniqueness of 
$\mathcal{L}_S\phi \in {L^1} (M)$ is equivalent by the Radon-Nikodym
Theorem to the absolute continuity (wrt.\ $\mu$) of the
measure associated to the functional
$\psi \mapsto \int_{M} d\mu^{N} \psi\circ S \, \phi$,
i.e.\ for all measurable $A\in M$, $\mu (A)=0$
implies $\mu (S^{-1}(A))=0$. This condition is called
\emph{nonsingularity} of $S$.

\vspace{0.5cm}

The normalized Lebesgue measure $\mu$ on $S^1$ is given by
$d\mu (z)=\frac{dz}{2\pi\imath}\frac{1}{z}$ (this lifts wrt.\ the
map $t\rightarrow e^{\imath t}$ to the normalized Lebesgue measure
$\frac{dt}{2\pi}$ on $[0,2\pi)$)
and the product measure $\mu^N$ on $(S^1)^N$ is given by

\beq
d\mu^N ({\bf z})=\frac{d{\bf z}}{(2\pi\imath)^N}\frac{1}{{\bf z}}
\stackrel{\mathrm{def}}{=}\frac{dz_1}{2\pi\imath} \cdots
\frac{dz_N}{2\pi\imath} \frac{1}{z_1}\cdots\frac{1}{z_N}
\eeq


\begin{proposition}

In the above setting the Perron-Frobenius-Operator can be written in
the following way: 
\la{proppfo}
\end{proposition}

\begin{equation}
\mathcal{L}_{S^\e} \phi (w) =\int_{\Gamma^N} \frac{d{\bf z}}{(2\pi\imath)^N}  
\phi (z) \prod_{k=1}^{N} \left(\frac{1}{S^{\e}_k({\bf z})-w_k}
\frac{S^{\e}_k({\bf z})}{z_k}\right)
\la{formreppfo}
\end{equation}

\emph{where $\Gamma =\Gamma_{+}\cup\Gamma_{-}$ is the 
positively-oriented boundary of $A_{\delta}$.}


\section{Further Remarks on the Infinite-Dimen/-sional System}
\la{sectionfurtherremarksontheinfinitedimensionalsystem}

The subspace of functions that depend only on finitely many variables
is dense in $\left( \mathcal{C}(M), \|\cdot\|_\infty \right)$, and each such
function (say depending on ${\bf z}_\Lambda$ only) can be uniformly
approximated by (the restriction of)
functions in $\mathcal{H}(A_\delta^\Lambda )$.
The dual space of $\mathcal{C}(M)$ is $rca(M)$ (see e.g.\
\cite{dunfordschwartz}),
the space of bounded, regular,
countably additive, real-valued set functions on $(M, \mathcal{B} )$
where $\mathcal{B}$ is the Borel $\sigma$-algebra. The norm on $rca (M)$
is the total variation.
For given $\vth ,\Lambda$ we consider \emph{rca} measures with marginals
${\phi_\Lambda}_{\mid (S^1)^\Lambda}$ over $(S^1)^\Lambda$
(restriction of $\phi_\Lambda$ to $(S^1)^\Lambda$)
s.t. $\phi = (\phi_\Lambda )_{\Lambda\in \mathcal{F}}
\in \mathcal{H}_{\vth}$.
We remark that not every $\phi\in\mathcal{H}_{\vth}$ with real-valued
marginals ${\phi_\Lambda}_{\mid (S^1)^\Lambda}$ 
corresponds to an element in $rca(M)$ 
because its variation might not be bounded
as 
$\int_\Lambda d \mu^\Lambda |\phi_\Lambda |$ might be unbounded
in $\Lambda$.
So we define for 
$\phi \in \mathcal{H}$

\beq
\|\phi\|_{var} \stackrel{\mathrm{def}}{=}
\lim_{\Lambda \to \Z^d} \int_{(S^1)^\Lambda} d \mu^\Lambda |\phi_\Lambda |.
\eeq

We set $\mathcal{H}^{bv} \stackrel{\mathrm{def}}{=}
\{ \phi \in \mathcal{H} : \|\phi\|_{var} < \infty \}$ and
$\mathcal{H}^{bv}_\vth \stackrel{\mathrm{def}}{=}
\mathcal{H}^{bv} \cap \mathcal{H}_\vth$. In particular all real-analytic 
and non-negative
$\phi\in\mathcal{H}$, i.e.\ ${\phi_\Lambda}_{\mid (S^1)^\Lambda} 
\geq 0$ for all 
$\Lambda\in\mathcal{F}$, belong to this space.

We can view every $\phi \in \mathcal{H}^{bv}$ as an element of
$rca (M)$: For $g \in \mathcal{C} (M)$ the net 
$(g_\Lambda )_{\Lambda \in \mathcal{F}}$ given by
$g_\Lambda \stackrel{\mathrm{def}}{=} \pi_\Lambda (g)$ converges
uniformly to $g$. We set
\beq
\phi (g) \stackrel{\mathrm{def}}{=}
\lim_{\Lambda \to \Z^d}
\int_{(S^1)^\Lambda} d \mu^\Lambda g_\Lambda \phi_\Lambda 
\eeq
The limit exists because for $\Lambda_1 \subset \Lambda_2$

\ba
\lefteqn{
\left| \int_{(S^1)^{\Lambda_1}} d \mu^{\Lambda_1} g_{\Lambda_1} 
\phi_{\Lambda_1}
-\int_{(S^1)^{\Lambda_2}} d \mu^{\Lambda_2} g_{\Lambda_2} 
\phi_{\Lambda_2} \right| }\\
& = & \left|\int_{(S^1)^{\Lambda_2}} d \mu^{\Lambda_2} 
(g_{\Lambda_1}-g_{\Lambda_2}) \phi_{\Lambda_2} \right| \nn\\
& \leq & \|g_{\Lambda_1}-g_{\Lambda_2}\|_{(S^1)^{\Lambda_2}}
\|\phi\|_{var}
\nn
\ea

gets arbitrarily small as $\Lambda_1 \rightarrow \Z^d$, i.e.\ the net 
has the Cauchy property.

We further see

\ba
\|\phi\|_{var} & = & 
\sup_{\Lambda \in \mathcal{F}}
\int_{(S^1)^\Lambda} d \mu^\Lambda | \phi_\Lambda | \\
& = & \sup_{\Lambda \in \mathcal{F}}
\sup_{g \in \mathcal{C}((S^1)^\Lambda) \atop
\| g \|_\infty \leq 1}
\int_{(S^1)^\Lambda} d \mu^\Lambda g \, \phi_\Lambda \nn\\
& = & \sup_{g \in \mathcal{C}(M) \atop 
\| g \|_\infty \leq 1}
| \phi (g)| \nn
\ea

so $\|\phi\|_{var}$ is in fact the total variation (the operator-norm,
cf.\ \cite{dunfordschwartz})
of the corresponding linear functional on $\mathcal{C} (M)$.

Let $\mathcal{H}(\mathcal{F})\stackrel{\mathrm{def}}{=}
\bigcup_{\Lambda\in\mathcal{F}}
H(A_\delta^\Lambda)$, the subspace of functions depending on only
finitely many variables.
We define the product 
$g^1 \phi \in \mathcal{H}_\vth$
of $g \in \mathcal{H}(A_\delta^{\Lambda_1})$ and
$\phi \in \mathcal{H}_\vth(\mathcal{F})$
by

\beq
(g^1 \phi )_\Lambda \stackrel{\mathrm{def}}{=}
\pi_{\Lambda} (g^1
\phi_{\Lambda_1 \cup \Lambda})
\la{defprod}
\eeq


\begin{lemma} If $g^1 \in H(A_\delta^{\Lambda_1})$, 
$g^2 \in H(A_\delta^{\Lambda_2})$,
$ g \in \mathcal{C}(M)$ and
$\phi \in \mathcal{H}_\vth$ the following holds
\la{lemmamult}
\end{lemma}

{\itshape
\begin{enumerate}
\item
The product in (\ref{defprod}) is well-defined and
$\|g^1 \phi \|_\vth \leq 
\| g^1\|_{\Lambda_1} \vth^{-|\Lambda_1 |} \|\phi\|_\vth$
\item
$(g^1 g^2) \phi = g^1 (g^2 \phi )$
\item
$g^2$ is also an element of
$\mathcal{H}_\vth$ and the product $g^1 g^2$ as defined in (\ref{defprod}) 
is the same as the usual product between functions on $M$.
\item
$(g^1 \phi )(g) = \phi (g^1 g)$ \quad where 
$(g^1 \phi )$ and $\phi$ act as functionals.
\item
$\mathcal{H}_\vth^{bv}$ is also a module over the ring
$\mathcal{H}(\mathcal{F})$.
\end{enumerate}
}


\section{Expansion of the Perron-Frobenius Operator}
\la{expansionoftheperronfrobeniusoperator}

We split the integral kernel of the Perron-Frobenius operator
for a finite-dimensional system.
Recall $S_p: M \to (S^1)^{\{ p\}}, \quad
S_p({\bf z}) = f_p \circ T_p^\e ({\bf z})$
with
$T^{\e}_p ({\bf z})= z_p \exp \left( 2\pi\imath\e 
\sum_{k=1}^{\infty} g_{p,k}({\bf z})\right)
=z_p \prod_{k+1}^{\infty} \exp (2\pi\imath\e g_{p,k}({\bf z}))$.

If we consider only finite range interaction, say up to distance $l$,
we have 

\beq
T^{\e}_{p,l} ({\bf z}) \stackrel{\mathrm{def}}{=}
z_p \exp (2\pi\imath\e \sum_{k=1}^l g_{p,k}({\bf z}))
\eeq

For a finite-dimensional system (say on $(S^1)^{\Lambda_2})$ with fixed 
boundary conditions we have a special representation  of
$\mathcal{L}_{F^{\Lambda_2}\circ T^{\Lambda_2 ,\e}}$
in terms of the integral kernel (Proposition \ref{proppfo}).


\begin{proposition}
For the factors in the integral kernel in (\ref{formreppfo}) 
we have the following splitting :
\la{propsplitting}
\end{proposition}

\ba
\lefteqn{
\frac{1}{f_p\circ T_p^{\e}({\bf z})-w_p}
\frac{f_p\circ T_p^{\e}({\bf z})}{z_p}
=
\frac{1}{f_p(z_p)-w_p}
\frac{f_p(z_p)}{z_p} } 
\la{formsplit}\\
& & +
\frac{w_p}{z_p}
\sum_{k=1}^{\infty}
\frac{f_p \circ T^\e_{p,k-1}({\bf z})- f_p\circ T^{\e}_{p,k}({\bf z})}
{\left( f_p \circ T^\e_{p,k-1} ({\bf z})-w_p\right) 
\left(f_p \circ T^\e_{p,k}({\bf z})-w_p\right)}
\nn
\ea

\emph{The sum in the right hand side converges uniformly in
${\bf z}\in \Gamma^N$ and $w_p \in A_\delta$.}


\subsection{Unperturbed Operator}
\la{subsectionunperturbedoperator}

The first summand in (\ref{formsplit})
is just the one which appears in the uncoupled system
(i.e.\ $T^{\e =0}=\id$) and 
in this case each lattice site can be considered separately.
We denote by
$\mathcal{L}_{f_p}$
the restriction of the Perron-Frobenius operator 
to the Banach space of functions on $S^1$ that extend 
continuously on the closed annulus $A_{\delta}$ and  holomorphically
on the interior $A_{\delta}$. $\|\cdot\|_{A_\delta}$ denotes the
supremum over $A_\delta$. The operator

\[ \mathcal{L}_{f_p}: \left(\mathcal{H}(A_{\delta}),
\|\cdot\|_{A_\delta}\right)
\rightarrow
\left(\mathcal{H}(A_{\delta}), \|\cdot\|_{A_\delta} \right)\]

has 1 as simple eigenvalue and the rest of 
its spectrum is contained in a disc around 0 of radius strictly
smaller than 1.  
It splits into

\beq
\mathcal{L}_{f_p}= Q_p +R_p
\eeq

with

\beq
R_p Q_p=Q_p R_p=0
\la{formcommute}
\eeq

and
\beq
\left\| R_p^n\right\|_{
L(\mathcal{H}(A_{\delta}), \|\cdot\|_{A_\delta})}\leq c_r\eta^n
\la{formrdecay}
\eeq
with $c_r>0$ , $0<\eta <1$. For proofs of these statements see e.g.
\cite{bricmontkupiainen1}.

$Q_p$ is the projection onto the one-dimensional eigenspace spanned by
$h_p\in \mathcal{H}(A_{\delta})$, whose restriction 
to $S^1$ is positive and has integral
$\int_{S^1}d\mu \, h_p = 1$.

We assume in condition {\bf I} regarding the family $(f_p)_{p\in \Z^d}$
that $\| h_p\|_\ad \leq c_h$ and that the exponential bound in
(\ref{formrdecay}) holds uniformly
in $p$. This is the case for example if the $f_p$ are uniformly close to
each other as is shown using analytic perturbation theory.

$\mathcal{L}_{f_p}$ preserves
the integral and so does $Q_p$ because of (\ref{formcommute})
and (\ref{formrdecay}). 
$\Gamma_+$ is homologous to $S^1$.
So we can write $Q_p$ as

\ba
Q_p g(w) & = & h_p(w)\int_{S^1}d\mu \, g \\
& = & h_p(w)\int_{\Gamma_+}\frac{dz}{2\pi\imath}\frac{1}{z} g(z)\\
& = & \int_{\Gamma}\frac{dz}{2\pi\imath}\frac{1}{z} h_p(w,z)g(z)
\ea

where we have used that $g$ is holomorphic in $A_\delta$ and
defined:

\beq
h_p(w_p,z_p)\stackrel{\mathrm{def}}{=}
\left\{
\begin{array} {c@{\quad \mbox{for} \quad}l}
h_p(w_p) &  z_p\in\Gamma_+\\
0        &  z_p\in\Gamma_-
\end{array}
\right.
\eeq

The idempotency $Q_p^2=Q_p$ reads in the integral representation

\beq
\int_{\Gamma} \frac{dz^2}{2\pi\imath} \frac{1}{z^2}
\int_{\Gamma} \frac{dz^1}{2\pi\imath} \frac{1}{z^1}
h_p(w,z^2) h_p(z^2,z^1) g(z^1)
=\int_{\Gamma} \frac{dz^1}{2\pi\imath} \frac{1}{z^1}
h_p(w,z^1)g(z^1)
\eeq

According to Proposition \ref{proppfo} the operator $R_p$ can be written

\beq
R_p g(w)=\int_{\Gamma} \frac{dz}{2\pi\imath} \frac{1}{z}
r_p(w,z) g(z)
\eeq

with
\beq
r_p(w,z)=\frac{1}{f_p(z)-w}
\frac{f_p(z)}{z} - h_p(w,z).
\eeq

Then equation
(\ref{formcommute}) reads in the integral representation

\ba
\int_{\Gamma} \frac{dz_p^2}{2\pi\imath} \frac{1}{z_p^2}
\int_{S^1} \frac{dz_p^1}{2\pi\imath} \frac{1}{z_p^1} \,
r_p(w_p,z_p^2) h_p(z_p^2) g(z^1) 
& = &
0,
\la{formzerohr}\\
\int_{S^1} \frac{dz_p^2}{2\pi\imath} \frac{1}{z_p^2}
\int_{\Gamma} \frac{dz_p^1}{2\pi\imath} \frac{1}{z_p^1} \,
r_p(z_p^2,z_p^1) g(z^1) 
& = &
0
\la{formzerorh}
\ea


\subsection{Perturbed Operator}

In view of (\ref{formsplit}) we set

\beq
\beta_{p,k}(w_p,{\bf z})\stackrel{\mathrm{def}}{=}
\frac{w_p}{z_p}
\frac{f_p\circ T_{p,k-1}^{\e}({\bf z}) -
f_p\circ T_{p,k}^{\e}({\bf z})}
{(f_p\circ T_{p,k-1}^{\e}({\bf z})-w_p)
(f_p\circ T_{p,k}^{\e}({\bf z})-w_p)} \quad .
\la{defbetak}
\eeq

This  corresponds to the difference between the operators for systems
with interaction of finite-range of order
$k$ and $k-1$, respectively. We have the estimate

\ba
\lefteqn{
| \beta_{p,k}(w_p,{\bf z})|}
\la{formestimatebetak}\\
& \leq &
\left|\frac{w_p}{z_p}\right|
|f_p\circ T_{p,k-1}^{\e}({\bf z})-w_p|^{-1}
|f_p\circ T_{p,k}^{\e}({\bf z})-w_p|^{-1}
\nn\\
& & \,\times
|f_p\circ T_{p,k-1}^{\e}({\bf z}) -f_p\circ T_{p,k}^{\e}({\bf z})| \nn\\
& \leq &
\frac{1+\delta}{1-\delta} |c_7\lambda -1|^{-1} |c_7\lambda -1|^{-1}
\| f_p'\|_{\{ p\}} c \e \exp (-c_2 k^d) \nn\\
& \leq &
c_8 \e \exp (-c_2 k^d)
\nn
\ea

uniformly in $p\in \Z^d$, $w_p \in A_\delta$, ${\bf z}\in M$.


\subsection{Time N Step}
\la{subsectionnntimeNstep}
Now we want to estimate the norm of (\ref{formpfoinf2}) or equivalently 
that of

\beq
\pi_{\Lambda_1} \circ \mathcal{L}^N_{F^{\Lambda_2} \circ 
T^{\Lambda_2 ,\e}} : 
\left(\mathcal{H}_{\Lambda_2 ,\vth} , 
\|\cdot\|_{\Lambda_2 ,\vth} \right)
\rightarrow
\left(\mathcal{H}_{\Lambda_1 ,\vth} , \|\cdot\|_{\Lambda_1 ,\vth} \right)
\la{formpfo21}
\eeq

\ba
\mathcal{L}^N_{F^{\Lambda_2}
\circ T^{\Lambda_2} ,\e}\phi ({\bf z}^0)
& =
\int_{\Gamma^{\Lambda_2}} \frac{d{\bf z}^{-1}}{(2\pi\imath)^{|\Lambda_2 |}}  
\cdots
\int_{\Gamma^{\Lambda}_2} \frac{d{\bf z}^{-N}}{(2\pi\imath)^{|\Lambda_2 |}} 
\prod_{t=-N}^{-1}
\prod_{p\in\Lambda_2}
\la{formintegralrepresentation}\\
& \times \left((h_p(z_p^{t+1},z_p^t) 
+ r_p(z_p^{t+1},z_p^t) +\sum_{k=1}^{\infty}
\beta_{p,k} (z_p^{t+1},{\bf z}^t)
\right)
\nn
\ea

Distributing the product we get infinitely many summands. In each factor
there is for each 
$-N\leq m\leq -1$, $p\in \Lambda_2$ a choice between
$h_p$, $r_p$ and $\beta_{p,k}$ $(1\leq k <\infty )$ 
and we can interpret such a choice 
graphically as a \emph{configuration} as follows 
(cf.\ \cite{bricmontkupiainen1, maesmoffaert}):

On $\Lambda_2 \times\{ -N,\ldots ,0\}$ we represent
\begin{itemize}
\item $h_p\left( z_p^{t+1},z_p^t\right)\quad$ by an \emph{h-line} from
$(p,t)$ to $(p,t+1)$

\item $r_p\left( z_p^{t+1},z_p^t\right)\quad$ by an \emph{r-line} from
$(p,t)$ to $(p,t+1)$


\begin{figure}[h]
\setlength{\unitlength}{0.5cm}
\begin{center}

\begin{picture}(0,5)

\put(-8,0){
\begin{picture}(0,5)

\put(0,0){\circle*{0.5}}
\put(0,4){\circle*{0.5}}

\put(0,0){\line(0,1){4}}
\put(0.5,3.8){$(p,t)$}
\put(0.5,-0.2){$(p,t+1)$}
\put(0.5,1.8){$h_p(z^{t+1}_p,z^{t}_p)$}
\end{picture}}

\put(4,0){
\begin{picture}(15,10)

\put(0,0){\circle*{0.5}}
\put(0,4){\circle*{0.5}}

{\thicklines \put(0,0){\line(0,1){4}}}
{\thicklines \put(0.05,0){\line(0,1){4}}}
{\thicklines \put(-0.05,0){\line(0,1){4}}}

\put(0.5,3.8){$(p,t)$}

\put(0.5,-0.2){$(p,t+1)$}
\put(0.5,1.8){$r_p(z^{t+1}_p,z^{t}_p)$}
\end{picture}}

\end{picture}
\end{center}

\caption{\la{figh-linerline}h-line and r-line}
\end{figure}

\item $\beta_{p,k}\left( z_p^{t+1},{\bf z}^t\right)\quad$ by a
\emph{k-triangle} (actually rather a cone or pyramid but in our pictures
for $d=1$ it is a triangle) with apex 
$(p,t+1)$ and base points $(q,t)$ with $\|p-q\|\leq k$.
(So some of the base points might not lie in
$\Lambda_2 \times \{ -N,\ldots , -1\}$ but all the apices lie in
$\Lambda_2 \times \{ -N+1,\ldots , 0\}$.)


\suppressfloats[t]
\begin{figure}[h!]

\setlength{\unitlength}{0.5cm}
\begin{center}

\begin{picture}(16,6)

\put(8,0){\circle*{.5}}
\put(0,4){\circle*{.5}}
\put(4,4){\circle*{.5}}
\put(8,4){\circle*{.5}}
\put(12,4){\circle*{.5}}
\put(16,4){\circle*{.5}}

\put(-1.3,4.5){$(p-2,t)$}
\put(2.8,4.5){$(p-1,t)$}
\put(7,4.5){$(p,t)$}
\put(10.9,4.5){$(p+1,t)$}
\put(14.9,4.5){$(p+2,t)$}
\put(6.8,-1.1){$(p,t+1)$}

\put(8,0){\line(2,1){8}}
\put(8,0){\line(-2,1){8}}
\put(0,4){\line(1,0){16}}

\put(6,2){$\beta_{p,2}( z_p^{t+1},{\bf z}^t)$}

\end{picture}
\end{center}

\caption{\la{figtriangle}2-triangle}
\end{figure}

\end{itemize}

Note that if $v(k)\stackrel{\mathrm{def}}{=} |B_k(0)|$
denotes the number of base points of a $k$-triangle, we have the estimate
$v(k)\leq (3k)^d$.

Each choice corresponds to a configuration and for each 
configuration $\mathcal{C}$ we have  an operator
$\mathcal{L_C}$. 
So we can write 
\beq
\mathcal{L}^N_{F^{\Lambda_2} \circ 
T^{\Lambda_2 ,\e}}  = \sum_{\mathcal{C}} \mathcal{L_C} 
\la{formpipfo1}
\eeq


Some of these summands are zero namely if
\begin{itemize}
\item a factor $h_p\left(z_p^{t+2},z_p^{t+1}\right)
r_p\left(z_p^{t+1},z_p^t\right)$
or
$r_p\left(z_p^{t+2},z_p^{t+1}\right)
h_p\left(z_p^{t+1},z_p^t\right)$
appears, but no factor
$\beta_{q,k}\left( z_q^{t+2},{\bf z}^{t+1}\right)$ with
$\|p-q\| \leq k$ (i.e.\ an h-line follows or is followed by an r-line
and at their common endpoint no triangle is attached with any of its
basepoints).
This follows since, by Fubini's Theorem, one can first perform the
$dz_p^{t+1} dz_p^t$-integration and get zero by (\ref{formzerohr})
or (\ref{formzerorh}). (Note
that no other terms depend on $z_p^{t+1}$ and the remaining factors
and integrations (up to time t+1) correspond to the function
$g(z^1)$ in (\ref{formzerohr}) or (\ref{formzerorh}).)


\begin{figure}[h]
\setlength{\unitlength}{0.5cm}
\begin{center}

\begin{picture}(0,8)

\put(-8,0){
\begin{picture}(0,8)

\put(0,0){\circle*{0.5}}
\put(0,4){\circle*{0.5}}
\put(0,8){\circle*{0.5}}
{\thicklines \put(0,4){\line(0,1){4}}}
{\thicklines \put(0.05,4){\line(0,1){4}}}
{\thicklines \put(-0.05,4){\line(0,1){4}}}
\put(0,0){\line(0,1){4}}
\put(0.5,3.8){$(p,t+1)$}
\put(0.5,7.7){$(p,t)$}
\put(0.5,5.8){$r_p(z^{t+1}_p,z^t_p)$}
\put(0.5,-0.2){$(p,t+2)$}
\put(0.5,1.8){$h_p(z^{t+2}_p,z^{t+1}_p)$}
\end{picture}}

\put(4,0){
\begin{picture}(15,10)

\put(0,0){\circle*{0.5}}
\put(0,4){\circle*{0.5}}
\put(0,8){\circle*{0.5}}
{\thicklines \put(0,0){\line(0,1){4}}}
{\thicklines \put(0.05,0){\line(0,1){4}}}
{\thicklines \put(-0.05,0){\line(0,1){4}}}
\put(0,4){\line(0,1){4}}
\put(0.5,3.8){$(p,t+1)$}
\put(0.5,7.7){$(p,t)$}
\put(0.5,5.8){$h_p(z^{t+1}_p,z^t_p)$}
\put(0.5,-0.2){$(p,t+2)$}
\put(0.5,1.8){$r_p(z^{t+2}_p,z^{t+1}_p)$}
\end{picture}}

\end{picture}
\end{center}

\caption{\la{figconsecutive}Consecutive r-line and h-line}
\end{figure}

\item if a term $h_p\left(z_p^{t+2},z_p^{t+1}\right)
\beta_{p,k}\left( z_p^{t+1},{\bf z}^t\right)$ appears but no
$\beta_{q,l}\left( z_q^{t+2},{\bf z}^{t+1}\right)$ with
$\|p-q\| \leq l$ (i.e.\ a triangle is followed by an h-line
and at their common endpoint (the apex of the triangle)
no other triangle is attached with any of its
basepoints).

\begin{eqnarray}
\beta_{p,k}\left( w_p,{\bf z}\right)
& = &
\frac{w_p}{z_p}
\frac{f_p\circ T_{p,k-1}^{\e}({\bf z}) -f_p\circ T_{p,k}^{\e}({\bf z})
}{(f_p\circ T_{p,k-1}^{\e}({\bf z}) - w_p)
(f_p\circ T_{p,k}^{\e}({\bf z}) - w_p)} \\
& = &\frac{w_p}{z_p}
\left[ \frac{1}{f_p\circ T_{p,k}^{\e}({\bf z})-w_p}
-
\frac{1}{f_p\circ T_{p,k-1}^{\e}({\bf z})-w_p}
\right]
\nn
\end{eqnarray}

By the Residue Theorem:
\beq
\int_{S^1}\frac{dw_p}{2\pi\imath}\frac{1}{w_p}
\beta_{p,k}\left( w_p,{\bf z}\right)
= 0
\la{formzerobetah}
\eeq

because the poles at
$ w_p = f_p\circ T_{p,k}^{\e}({\bf z})$ and 
$ w_p = f_p\circ T_{p,k-1}^{\e}({\bf z})$
(with ${\bf z}\in\Gamma^N$, in particular $z_p\in\Gamma_{+}$
or $\Gamma_{-}$)
both lie either outside $\Gamma_+$ or inside $\Gamma_-$
as $f_p$ is expanding and $T_{p,k}^{\e}$ close to $T_{p,k-1}^{\e}$
and the two summands have residue
$\frac{-1}{z_p}$ and $\frac{1}{z_p}$, respectively.

This identity is a consequence of the fact that 
$\beta_{p,k}$ is the kernel of a 
difference between two transfer operators (for the systems with interaction
of range $k$ and $k-1$) both preserving the integral. So the range
of this operator consists of functions with integral zero
and these are annihilated by the operator corresponding to $h_p$.


\begin{figure}[h]

\setlength{\unitlength}{0.5cm}

\begin{center}
\begin{picture}(0,9)

\put(0,4){\circle*{.5}}
\put(-8,8){\circle*{.5}}
\put(-4,8){\circle*{.5}}
\put(0,8){\circle*{.5}}
\put(4,8){\circle*{.5}}
\put(8,8){\circle*{.5}}

\put(0,0){\circle*{.5}}

\put(-1,8.5){$(p,t)$}
\put(0.4,3.5){$(p,t+1)$}

\put(-1.2,-1.5){$(p,t+2)$}

\put(0,4){\line(2,1){8}}
\put(0,4){\line(-2,1){8}}
\put(-8,8){\line(1,0){16}}
\put(0,0){\line(0,1){4}}

\put(-2,6){$\beta_{p,2}( z_p^{t+1},{\bf z}^t)$}

\put(0.5,1.8){$h_p(z^{t+2}_p,z^{t+1}_p)$}

\end{picture}
\end{center}

\caption{\la{figcombinationtrianglehline}Combination 2-triangle and h-line}
\end{figure}

\end{itemize}

Furthermore we note that in 
\beq
\pi_{\Lambda_1} \circ \mathcal{L}^N_{F^{\Lambda_2} \circ 
T^{\Lambda_2 ,\e}} =
\sum_{\mathcal{C}} \pi_{\Lambda_1} \circ \mathcal{L_C}
\la{formpipfo2}
\eeq
we get $\pi_{\Lambda_1} \circ \mathcal{L_C}=0$ unless $\mathcal{C}$
ends with h-lines in all points of 
$\left( \Lambda_2 \setminus \Lambda_1 \right)\times \{ 0 \}$
because of (\ref{formzerorh}), (\ref{formzerobetah}) 
and the fact that $\pi_{\Lambda_1} $ 
means integration
over $(S^1)^{\Lambda_2 \setminus \Lambda_1 }$.
So we just have to sum over non-zero configurations that end (at time 0)
with r-lines or triangles at most in $\Lambda_1 \times \{ 0\}$.
Let $\mathcal{C}$ be a configuration with exactly $n_r$ r-lines and
$n_{\beta ,k}$ $k$-triangles for $0\leq k < \infty$
(so the set of triangles is given by
$n_\beta \stackrel{\mathrm{def}}{=}
(n_{\beta ,1}, n_{\beta ,2}, \ldots )$ with
$|n_\beta |\stackrel{\mathrm{def}}{=} 
\sum_{k=1}^{\infty} n_{\beta ,k} < \infty$).

\vspace{0.3cm}

We have to find an upper bound for the norm of each 
$\mathcal{L_C}$. We do so by collecting r- and h-lines into
chains and estimating the contributions of integrating the factors
corresponding to these parts of the configuration.

\vspace{0.3cm}


{\bf Definition}
\la{def1}
A sequence of lines from $(p,t)$ to
$(p,t+1)$, $\ldots$, $(p,t+k-1)$ to $(p,t+k)$ with $p\in \Lambda_2$
and $-N \leq t \leq t+k \leq 0$ such that to the points
$(p,t+1)\ldots (p,t+k-1)$ no triangles are attached is called an
\emph{h-chain of length k}. If such an h-chain is not contained in a
longer one it is called a \emph{maximal h-chain}. Then $(p,t)$ 
and $(p,t+k)$ are denoted its \emph{endpoints}.
The definitions of \emph{r-chain} etc. are analogous.
Furthermore let $\tilde{\Lambda}_{\mathcal{C}}$ be the set of points
$p\in \Lambda_2$
that appear as the $\Z^d$-coordinate
of a base point $(p,t)$ of a triangle in $\mathcal{C}$
and $\Lambda_{\mathcal{C}}$ the set of those points $p \in \Z^d$
that appear as the $\Z^d$-coordinate of an apex $(p,t)$ that does not lie
above any other triangle.
$\Lambda_r$ is the set of $r\in \Z^d\setminus
\tilde{\Lambda}_{\mathcal{C}}$ that appear as the $\Z^d$ coordinate of an 
r-line (this implies that there is an r-chain from time $-N$ to time $0$).
$\Lambda (\mathcal{C}) \stackrel{\mathrm{def}}{=}
\tilde{\Lambda}_{\mathcal{C}} \cup \Lambda_r$.

In Figure \ref{figexampleconfiguration} 
there are for example maximal r-chains from
$(1, -3)$ to $(1,0)$ or from $(2, -3)$ to $(2, -2)$.
$\Lambda_2 = \{ 1,\ldots ,8 \}$,
$\tilde{\Lambda}_{\mathcal{C}}=\{ 2,\ldots , 7\}$,
$\Lambda_{\mathcal{C}}= \{ 4\}$
and $\Lambda_r = \{ 1\}$.


\begin{figure}[h]
\begin{center}
\begin{minipage}[b]{\linewidth}
\centering 
\epsfig{file=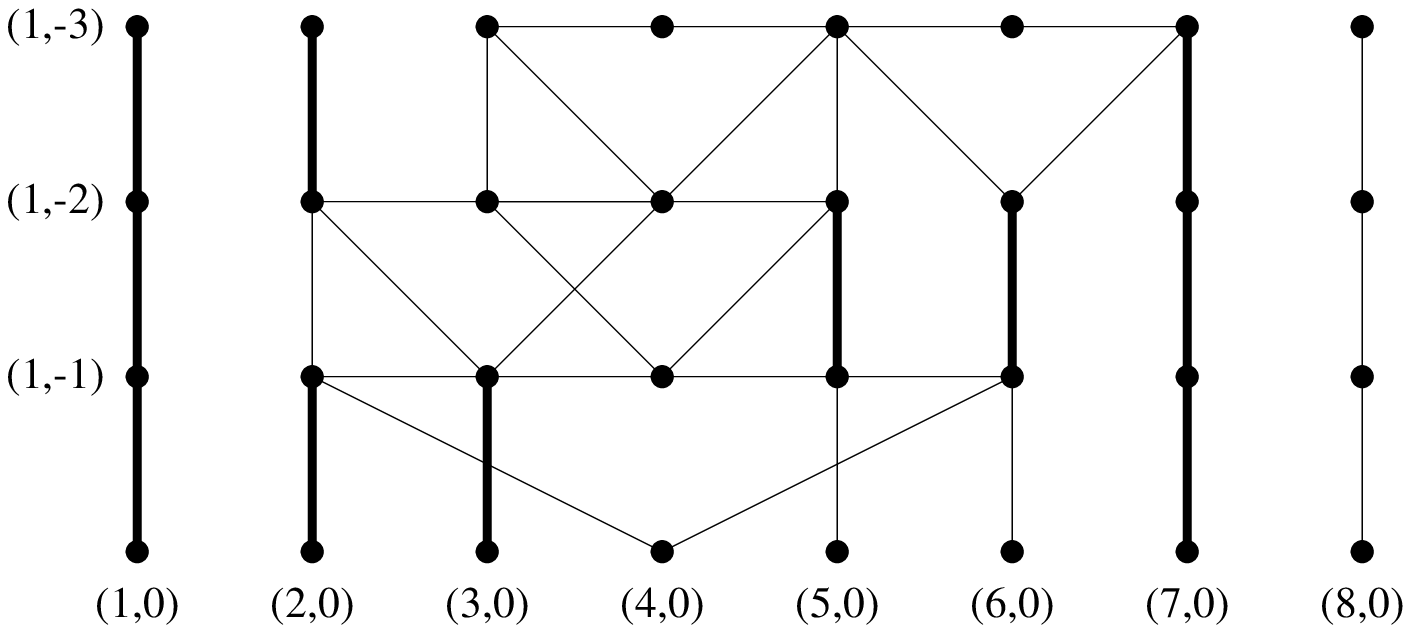,width=\linewidth}
\end{minipage}
\caption{\la{figexampleconfiguration}Example for a configuration}
\end{center}
\end{figure}

As each $k$-triangle has $v(k) \leq (3k)^d$ base points we have
\beq
\tilde{\Lambda}_{\mathcal{C}} \leq
\sum_{k=1}^{\infty} (3k)^d n_{\beta ,k}
\la{formestimatelambdac}
\eeq
 
To get the estimate for (\ref{formpfo21}) 
we proceed in the following order:

\begin{enumerate}
\item 
We integrate in $\left| \pi_{\Lambda_1} \circ\mathcal{L_C}
\phi\left({\bf z}_{\Lambda_1}^0\right)\right|$ over all
$dz_p^t$ for which a factor\\
$r_p(z_p^{t+1},z_p^t)$
appears. For each maximal r-chain of length $l$ we 
get according to (\ref{formrdecay})
a factor not greater than $c_r\eta^l$.

\item 
For each maximal h-chain starting at $(p,t)$ and ending at
$(p,t+l)$ we perform the integration

\beq
\int_{\Gamma} \frac{d z_p^{t+l-1}}{2\pi\imath} \cdots
\int_{\Gamma} \frac{d z_p^{t+1}}{2\pi\imath}\,
h_p(z_p^{t+l},z_p^{t+l-1})\cdots h_p(z_p^{t+1},z_p^t)
=h_p(z_p^{t+l})
\eeq

\item We perform the integration corresponding to $\pi_{\Lambda_1}$

\beq
\prod_{p\in\Lambda_2 \setminus \Lambda_1 }
\int_{S^1}
\frac{d z_p^0}{2\pi\imath}\, h_p(z_p^0) = 1
\eeq

\item 
In the remaining integral we estimate uniformly
$|\beta_{p,k}(z_p^{t+1},{\bf z}^t)|$ by (\ref{formestimatebetak})
and each (from step 2 and 3 remaining) factor $h_p(z_p^t)$ by
$\|h_p\|_{A_\delta}\leq c_h$
and $|\phi ({\bf z}^{-N})|$ by
$\|\phi_{\tilde{\Lambda}_{\mathcal{C}} \cup \Lambda_r} 
\|_{A_\delta^{\tilde{\Lambda}_{\mathcal{C}}\cup \Lambda_r}}$
(cf.\ remark below).

\end{enumerate}


{\bf Remark}
For all points $q \not\in \tilde{\Lambda}_{\mathcal{C}}
\cup \Lambda_r$
we must have h-chains in $\mathcal{C}$ from $(q,-N)$ to
$(q,0)$. Therefore we have

\beq\pi_{\Lambda_1} \circ \mathcal{L_C} \phi_{\Lambda_2}
({\bf z}^0_{\Lambda_1})=
\pi_{\Lambda_1} \circ \mathcal{L_C} \phi_{\tilde{\Lambda}_{\mathcal{C}}
\cup \Lambda_r}
({\bf z}_{\Lambda_1}^0)
\eeq

where on the righthandside we use the same notation `$ \mathcal{L_C}$'
for the operator on
$H_{A_\delta^{\tilde{\Lambda}_{\mathcal{C}}\cup \Lambda_r ,\vth}}$.

\vspace{0.5cm}
So if $\tilde{n}_r$ denotes the number of maximal r-chains and
$\tilde{n}_h$ the number of maximal h-chains having spatial 
coordinates in $\tilde{\Lambda}_{\mathcal{C}} \cup \Lambda_1$ (for otherwise
they are `integrated away' giving a factor of 1)
we get

\ba
\lefteqn{
\left\| \pi_{\Lambda_1} \circ\mathcal{L_C} \phi
\right\|_{\Lambda_1}}
\la{formestimatepi1lc}\\
& \leq & 
(c_1 \e )^{|n_\beta |} 
\exp \left( -c_2 \sum_{k=1}^{\infty} k^d n_{\beta ,k} \right)
c_h^{\tilde{n}_h} c_r^{\tilde{n}_r}
\eta^{n_r}
\|\phi_{\tilde{\Lambda}_{\mathcal{C}}\cup\Lambda_r}
\|_{\tilde{\Lambda}_{\mathcal{C}}\cup \Lambda_r}
\nn
\ea

with

\ba
\|\phi_{\tilde{\Lambda}_{\mathcal{C}}\cup\Lambda_r}
\|_{\tilde{\Lambda}_{\mathcal{C}}\cup\Lambda_r}
& \leq &
\vth^{-|\Lambda_r|-\sum_{k=1}^{\infty} (3k)^d n_{\beta ,k}}
\|\phi\|_{\Lambda_2 ,\vth}
\la{formestimatephilambdaclambdar}\\
& \leq &
\vth^{-|\Lambda_r|} \prod_{k=1}^{\infty}\vth^{-(3k)^d n_{\beta ,k}}
\|\phi\|_{\Lambda_2 ,\vth}
\nn
\ea

for all $\Lambda_2 \in \mathcal{F}$.

\vspace{0.3cm}


\section{Operators for the Infinite-Dimensional\\ System}
\la{sectionoperatorsfortheinfinitedimensionalsystem}

Estimates (\ref{formestimatepi1lc}) and 
(\ref{formestimatephilambdaclambdar}) bound the particular summands 
in an expansion like (\ref{formintegralrepresentation}).
We see that triangles and maximal r-chains in a configuration
$\mathcal{C}$ lead to small factors on the right hand side
of (\ref{formestimatepi1lc}). (A maximal r-chain consisting of $n$
r-lines contributes a factor $c_r \eta^n$. The factor $c_r$ is greater
than 1 in general. But either it will be compensated for by a small factor
due to a triangle e.g.\ as in (\ref{formestimate})
or $n$ will be large, cf.\ e.g.\ (\ref{formestimatecretaN})).
This motivates the following definition of the length of a configuration.
The length gives rise to a lower bound for the number of triangles or
r-lines, i.e.\ a long configuration will lead to a small contribution
in the total sum in (\ref{formintegralrepresentation}).


{\bf Definition} The \emph{length}, $\mbox{length} (\mathcal{C})$, of 
a configuration $\mathcal{C}$ (that we got in an expansion like 
(\ref{formpipfo1}))
is the maximal difference $0-t$ such that there are points
$(p,t)$ and $(q,0)$ being end-points of r-lines or base points or apices 
of triangles. (Note that if there are any triangles or r-lines, there
is also a triangle or r-line ending at $\Lambda\times\{ 0\}$.)
If there are no triangles or r-lines in $\mathcal{C}$
its length is zero.

We identify two non-zero
configurations $\mathcal{C}_1$ and $\mathcal{C}_2$ 
if they agree in their triangles and r-lines (but might have
different $t_0, \Lambda_2$).
Then for a  configuration
$\mathcal{C}$ \quad $\mbox{length}(\mathcal{C})$,
$L(\mathcal{C})$, $\tilde{\Lambda}_{\mathcal{C}}$,
$\Lambda (\mathcal{C})$ (as in the definition on 
page \pageref{def1}) 
and the operator
$\pi_\Lambda \circ \mathcal{L_C}
\in L((\mathcal{H}(A_\delta^{\Lambda (\mathcal{C})}), 
\|\cdot\|_{\Lambda (\mathcal{C})}),
(\mathcal{H}(A_\delta^{\Lambda}), \|\cdot\|_{\Lambda}))$
are still well-defined.

For $\Lambda_1 \in \mathcal{F}$ we define $E(\Lambda_1 )$ 
as the set of all 
non-zero configurations $\mathcal{C}$ in some 
$\Lambda_2 \times\{ -t_0, \ldots , 0\}$ 
with
$\Lambda_1 \subset \Lambda_2 \in \mathcal{F}$,
$t_0\in N$ and $t_0> \mbox{length}(\mathcal{C})$,
and that end at time 0 with triangles or r-lines at most in
$\Lambda_1 \times\{ 0\}$.

Further we define $E_N(\Lambda_1 )$ as the set of
non-zero configurations $\mathcal{C}$ in
$\Lambda_2 \times\{ -N,\ldots , 0\}$ 
with
$\Lambda_1 \subset \Lambda_2 \in \mathcal{F}$ and
$\Lambda_{\mathcal{C}} \subseteq \Lambda_2$.

\vspace{0.3cm}

We define

\beq
\nu_\Lambda \stackrel{\mathrm{def}}{=}
\sum_{\mathcal{C}\in E(\Lambda )} \pi_\Lambda \circ \mathcal{L_C}
h_{\Lambda (\mathcal{C})}
\la{defnu}
\eeq

The convergence of this infinite sum and other properties of
$\nu$ are proved in the following proposition additional to Theorem
\ref{theoremexistencepfo}.


\begin{proposition}
Let $\vth$, $\e$, $c_2$, $N$ and $\Lambda_1$ as 
in Theorem \ref{theoremexistencepfo}.
\la{propexpfo}
\end{proposition}

{\itshape
\begin{enumerate}
\item
\beq
\pi_{\Lambda_1} \circ \mathcal{L}^N_{F \circ T^\e}
= 
\sum_{\mathcal{C}\in E_N (\Lambda_1)}
\pi_{\Lambda_1}\circ \mathcal{L_C}
\la{formpi1pfoN}
\eeq

\item
\beq
\left\|\mathcal{L}^N_{F \circ T^\e}
- \mathcal{L}^{N+1}_{F \circ T^\e}\right\|_{L
\left(\left(\mathcal{H}_{\vth} ,\|\cdot\|_{\vth}
\right)\right)}
\leq c_9 \tilde{\eta}^N
\la{formcauchycritpfo}
\eeq

\item
$\nu =(\nu_{\Lambda_1} )_{\Lambda_1 \in\mathcal{F}}\in
\mathcal{H}_{\vth}^{bv}$, $\mu (\nu )=1$ and $\nu \geq 0$
and $\nu$ satisfies (\ref{formestimatepfoNminusmu}) in Theorem
\ref{theoremexistencepfo}.

\item
For $N_1 , N_2 \in N$ the operator $\mathcal{L}_{F \circ T^\e}^{N_2}$
is defined on
$\mathcal{L}_{F \circ T^\e}^{N_1} (\mathcal{H}_\vth )
\subset \mathcal{H}_{\vth_{N_1}}$
and maps this space to 
$\mathcal{H}_{\vth_{N_1+N_2}}$
and

\beq
\mathcal{L}_{F \circ T^\e}^{N_2} \circ 
\mathcal{L}_{F \circ T^\e}^{N_1} =
\mathcal{L}^{N_1+N_2}_{F \circ T^\e}
\la{formsemigroup}
\eeq

\item
$\nu$ is the unique 
$\mathcal{L}_{F \circ T^\e}$-invariant element in 
$\mathcal{H}_{\vth}$ with $\mu (\nu ) = 1$

\item
For $g\in \mathcal{C} (M)$ and 
$\phi\in \mathcal{H}^{bv}_\vth$

\beq
\int_M d\mu \, g \circ S \: \phi
= \int_M d\mu \, g \mathcal{L}_{F\circ T^\e} \phi
\eeq

and in particular

\beq
\mu (\phi) = \mu (\mathcal{L}_{F\circ T^\e} \phi)
\eeq

For finite-range interaction all this also holds for 
$\phi\in \mathcal{H}^{bv}$.

\item
$\mathcal{L}_{F\circ T^\e}$ is non-negative, i.e.\ $\phi \geq 0$
implies $\mathcal{L}_{F\circ T^\e} \phi \geq 0$.

\item
For $\phi\in \mathcal{H}^{bv}_\vth$ we have the estimate  
$\|\mathcal{L}_{F\circ T^\e} \phi\|_{var} \leq \|\phi\|_{var}$.
For finite-range interaction all this also holds for 
$\phi\in \mathcal{H}^{bv}$.

\end{enumerate}}


\section{Decay of Correlations}
\la{sectiondecayofcorrelations}
We have found the unique $\nu \in \mathcal{H}_\vth$
with $\mu (\nu )=1$. This corresponds to a non-negative measure
on $(M, \mathcal{B})$ whose marginal on 
$(S^1)^\Lambda$ wrt.\ $\mu^\Lambda$ is given by (the restriction of)
$\nu_\Lambda$.
We need the following proposition about exponential decay of correlations
for $\nu$ for the proof of Theorem \ref{theoremdecay2}.


\begin{theorem}
For sufficiently small $\vth$ and $\e$, big $c_2$, finite
disjoint $\Lambda_1 , \Lambda_2$
and $f\in H(\adz )$
there are a $\kappa\in (0,1)$ and a $\tilde{\vth}\in (0,1)$ such that
\la{propositiondecay1}
\end{theorem}

\begin{enumerate}
\item
$\left\| \nu_{\Lambda_1 \cup \Lambda_2} - \nu_{\Lambda_1}
\nu_{\Lambda_2} \right\|_{\Lambda_1 \cup \Lambda_2 ,\vth}
\leq c_{10} \kappa^{\dist}$

\item
$\| \pi_{\Lambda_1}(f \nu ) - \nu (f) \nu_{\Lambda_1}\|_{\Lambda_1 ,\vth}
\leq
c_{11} \vth^{-|\Lambda_2|}\| f\|_{\Lambda_2 } 
\kappa^{\dist}$

\item
$\| \pi_{\Lambda_1}\circ\mathcal{L}^N_{F
\circ T^{\e}}
(f\nu )  - \nu (f) \nu_{\Lambda_1}\|_{\Lambda_1 ,\tvth}
\leq
c_{12} \vth^{-|\Lambda_2|} \| f\|_{\Lambda_2} 
\kappa^{\dist} \tilde{\eta}^N$

for every $N\geq 0$.
\end{enumerate}

{\bf Remark }As in Theorem \ref{theoremdecay2} we can
choose the rate of decay $\kappa$ first and then the other parameters.


\section{Proofs}
\la{sectionproofs}

{\bf Proof of Proposition \ref{propcd} }
We have 
a Cauchy estimate for the partial derivatives of the functions
$g_{p,k} : A_{{\delta}}^{B_k(p)} \rightarrow \C$ on a smaller polyannulus. 
Let $q\in B_k(p)$, Then

\begin{eqnarray}
{\left\| \frac{\partial}{\partial z_q} g_{p,k}
\right\|}_{A_{\delta}^{B_k(p)}}
& \leq &
\frac{1}{
| e^{\delta} -e^{\delta_1}|}
c_1 \exp (-c_2 k^d)\\
& = & c_{13} \exp (-c_2 k^d) \nn\\
\end{eqnarray}

Also note that $\frac{\partial}{\partial z_q} g_{p,k} =0$ for
$q \notin B_k(p)$.
Therefore 

\begin{eqnarray}
\left\|\frac{\partial}{\partial z_q} 
g_p \right\|_{A_{{\delta}}^{\Z^d}} & = & 
 \left\| \frac{\partial}{\partial z_q} \sum_{k=\| p-q\|}^{\infty} g_{p,k}
\right\|_{A_{{\delta}}^{\Z^d}} 
\la{formdgp}\\
 & \leq & c_{13} \sum_{k=\|p-q\|}^{\infty} \exp (-c_2 k^d)\nn\\
 & \leq & c_{13} \frac{1}{1-\exp (-c_2)} \exp \left(-c_2\| p-q\|^d\right)\nn\\
 & = & c_{14} \exp \left(-c_2\| p-q\|^d\right)
\nn
\end{eqnarray}

Now we consider everything in the lift given by
$ pr: 
\C_{\delta}^{\Lambda}\rightarrow A_{\delta}^{\Lambda}$,
$\left( \tilde{z_p} \right)_{p\in\Lambda}\mapsto 
\left( e^{\imath\tilde{z_p}}\right)_{p\in\Lambda}$,
where  $\C_{\delta} \stackrel{\mathrm{def}}{=} \left\{ 
w\in \C\mid\Im w \in [ -\delta, \delta ] \right\}.$

Then we have 
$\left(\widetilde{T^{\Lambda ,\e}} \left(\tilde{{\bf z}}\right)
\right)_p = \tilde{z}_p +2\pi\e \tilde{g_p}(\tilde{{\bf z}})$.
The function
$\tilde{g_p} (\tilde{\bf z})= g_p(pr(\tilde{{\bf z}}))$ 
satisfies the same
estimate (\ref{formgpk}) with a different constant $\tilde{c_1} $
for $\delta <\delta_1$ sufficiently small since $pr$ and 
its partial derivatives
are uniformly bounded on $\C_{\delta}^{\Lambda}$.

Then we have

\[ \left|  D \left(\widetilde{T^{\Lambda ,\e}} \left(\tilde{\bf z}\right)
  \right)_{p,q} - \delta_{p,q} \right|
\leq 2\pi\e \tilde{c_1}\exp \left( -{c_{13}} \| p-q\|^d \right)
\]

In particular the row sum norm (the operator-norm induced by the 
$l^{\infty}$-norm on $\C^{\Lambda}$) of
$ \left( D \widetilde{T^{\Lambda ,\e}} - \id\right)$ is smaller than 1 for
$\e < \e_0$ small enough. According to Lemma \ref{lemmamap} 
(cf.\ below and noting that
$\C_{\delta}$ is convex),
$\widetilde{T^{\Lambda ,\e}}$ is a biholomorphic map onto its image
and so is $T^{\Lambda ,\e}$.

Now fix $\delta < \delta_0$ according to the first part of the proof.
If ${\bf z} \in \partial A_{\delta}^{\Lambda}$ we have $z_p \in 
\partial A_{\delta}$
for at least one $p \in \Lambda$. From the  formula
$z_p^{'} \stackrel{\mathrm{def}}{=} T_p^{\Lambda , \e}({\bf z})
= z_p \exp \left( 2 \pi \imath \e g_p ({\bf z}) \right)$
and the assumption that $g_p$ is uniformly bounded on $A_{\delta_1}$
we see

\beq
\left| \ln |z'_p|\right| \geq \delta -c_{16}\e >c_{17}\delta
\la{formestimatecd}
\eeq

for $\e \leq \e_0< \frac{1-c_{17}}{c_{16}}\delta$.

Now assume $\emptyset \not= A_{c\delta}\setminus 
T^{\Lambda , \e}\left(A_{\delta}\right)\ni {\bf z}$. 
Let $s$ be the line-segment between {\bf z} and its nearest point {\bf w} 
on $\left( S^1\right)^{\Lambda}$ (wrt.\ the metric $d_{\Lambda}$).
For each point {\bf y} on $s$ the inequality 
$\ln d_{\Lambda}({\bf w},{\bf y})\leq 
\ln d_{\Lambda} ({\bf w},{\bf z}) \leq c_{17} \delta $
holds.

In particular there is a 
${\bf y}\in T^{\Lambda , \e}\left(\partial A_{\delta}^{\Lambda}\right)$ 
on $s$
with $\left| y_p\right| \leq  c_{17} \delta$ for all $p\in\Lambda$, but 
this contradicts the estimate (\ref{formestimatecd}) above.
\qed


\begin{lemma}
If $T: U\rightarrow \C^n$ is a holmorphic map on a convex
set $U\subset \C^n$ and satisfies the estimate
$\left\| DT( z) - \id\right\| \leq c_{18}<1$ 
then T is biholomorphic onto its image (in this lemma
the chosen norm on $\C^n$ and the corresponding operator norm 
are both denoted by $\|\cdot\|$).
\la{lemmamap}
\end{lemma}

{\bf Proof } T is locally biholomorphic by the Inverse Function Theorem.
So we only have to show injectivity. Let $z^0,z^1\in U$ with 
$T\left(z^0\right) = T\left(z^1\right)$ and 
$\gamma : [0,1]\rightarrow U$ , $\gamma (t)=z^o + t (z^1-z^0)$.
Then
\begin{eqnarray}
\left\| z^1-z^0\right\| & = & \left\| T\left( z^1\right) +z^1 - 
                           T\left( z^0\right) -z^0\right\|\nn\\
 & = & \left\| T\circ\gamma (1)+\gamma (1)-
       T\circ\gamma (0)+\gamma (0)\right\|\nn\\
 & = & \left\|\int_{0}^{1} \left( DT\left(\gamma (t)\right)-\id\right)
       \left( z^1-z^0\right) dt\right\| \nn\\
 & \leq & \left\| z^1-z^0\right\| \int_{0}^{1}
          \left\| DT\left(\gamma (t)\right)-\id\right\| dt \nn\\
 & \leq & \left\| z^1-z^0 \right\| c_{18}
\end{eqnarray}

which implies $z^1=z^0$.
\qed


{\bf Proof of Proposition \ref{propld} }
As $F$ acts on each coordinate separately by an $f_p$ we have (in view
on the chosen metric (\ref{definitionlambdametric})) to show the statement
just for the map f (we drop the index $p$), i.e.
the case $\Lambda$ containing just one element.

Consider the lift $\R_{\delta}\times \R \ni (r,\phi)\mapsto
r e^{\imath \phi}$ 
where $\R_{\delta} \stackrel{\mathrm{def}}{=} [1-\ln\delta , 1+\ln\delta ]$.
This defines (modulo $(0,2\pi )$ ) a  $(0,2\pi )$-periodic map
$\tilde{f}=\left(\tilde{f}_r ,\tilde{f}_{\phi}\right)$ 
via

$f \left( r e^{\imath \phi}\right) =
\tilde{f}_r (r,\phi ) e^{\imath \tilde{f}_{\phi}(r,\phi )}$.
On $\{ 1\}\times \R$ one has 
$\frac{\partial}{\partial r}\tilde{f}_r \geq\lambda_0$
and so because of periodicity and a compactness argument,
$\frac{\partial}{\partial r}\tilde{f}_{r} \geq\lambda$
on a thin ($0<\delta <\delta_0$ small)
strip $\R_{\delta}\times \R$.
It follows similarly, as in the proof of Proposition \ref{propcd}, that
$\tilde{f}\left( \R_{\delta}\times \R\right)\supset
\R_{\lambda\delta}\times \R$,
$\tilde{f}$ is diffeomorphic onto its image and each point in 
$\R_{\delta}\times \R$
has the same number of preimages (which is equal to
$\left(\tilde{f} (1,2\pi )-\tilde{f}(1,0)\right) / 2\pi$).
>From this the claim about $f$ follows.
\qed


{\bf Proof of Proposition \ref{proppfo} }
We substitute the expression (\ref{formreppfo}) into the right-hand side of 
equation (\ref{defpfo}) and get

\beq
\int_{T^N} \frac{d{\bf w}}{(2\pi\imath)^N}\frac{1}{{\bf w}}
\psi ({\bf w}) \int_{\Gamma^n} \frac{d{\bf z}}{(2\pi\imath)^N} 
\phi ({\bf z}) \prod_{k=1}^{N} \left(\frac{1}{S^\e_k({\bf z})-w_k}
\frac{S^\e_k({\bf z})}{z_k}\right) 
\eeq

As (\ref{defpfo}) is linear in $\psi$ we can assume (by using a continuous
partition of unity) that $\psi$ vanishes 
outside a small set $K\subset T^N$ having distinct preimages 
under $S^t$ (for all $0 \leq t \leq \e$) contained
in $K_{\alpha}=K_{\alpha_1}\times\cdots\times K_{\alpha_N}$ such that each
$K_{\alpha}$ is contained in a polydisc $D_{\alpha}=
D_{\alpha_1}\times\cdots\times D_{\alpha_N}$. These are
mutually disjoint and $S_\alpha^{t} \stackrel{\mathrm{def}}{=}
S_{| K_{\alpha}}^t$ is biholomorphic onto $K$ (for all $0 \leq t \leq \e$).
(To make this more precise we note that for $t=0$ the 
map $S^0$ is the product of maps $f_i$ ($1\leq i \leq N$)
and each $f_i$ gives rise to an $M_i$-fold covering map
of $A_\delta$. So locally we can index the disjoint preimages of $K$
under $S^0$ by $\alpha = (\alpha_1,\ldots ,\alpha_N)$ where
$1\leq \alpha_i \leq M_i$. If we take the set $K$ small enough this is 
still true under small ($0 \leq t \leq \e$) perturbations.)

For given ${\bf w}\in K$, index $\alpha$ as above, $k\in \{1,\ldots ,N\}$
and fixed $z_l\in K_{\alpha_l}$ $(l\neq k)$ the function
$z_k\mapsto (S^{\e}_{\alpha ,k}(z_1,\cdots ,z_k,\cdots ,z_N)-w_k)^{-1}$
has exactly one simple pole in $D_{\alpha_k}$ and is holomorphic in
$A_{\delta}^{\Lambda}$ away from this pole. Therefore we get the same
if we just integrate around these poles. 

\beq
=\int_{K} \frac{d{\bf w}}{(2\pi\imath)^N} \frac{1}{{\bf w}}
 \psi ({\bf w}) \sum_{\alpha}
\left( \prod_{k=1}^{N}
\int_{\partial D_{\alpha_k}} \frac{d z_k}{2\pi\imath} \right)
\phi ({\bf z}) \prod_{k=1}^N\frac{S^{\e}_{\alpha ,k}({\bf z})}{z_k}
\prod_{k=1}^N\frac{1}{S^{\e}_{\alpha ,k}({\bf z})-w_k}
\la{formintegration1}
\eeq

For each $\alpha$ we can write each of the inner integrals as an integral of
a differential form over the distinguished boundary
$b_0 D_{\alpha} \stackrel{\mathrm{def}}{=}
\partial D_{\alpha_1} \times \ldots \times \partial D_{\alpha_N}$,
parameterized by
$[ 0,1)^N \ni t \mapsto (e^{2\pi\imath t_1},\ldots ,
e^{2\pi\imath t_N})$, whence

\beq
\int_{b_0 D_{\alpha}} \phi ({\bf z})
\prod_{k=1}^N\frac{S^{\e}_{\alpha ,k}({\bf z})}{z_k}
\prod_{k=1}^N\frac{1}{S^{\e}_{\alpha ,k}({\bf z})-w_k}
dz_1 \wedge \ldots \wedge dz_N
\eeq

We want to split the singular factor into a product of single poles in
each variable. So we apply the transformation
${\bf u}=S_\e({\bf z})\stackrel{\mathrm{def}}{=}  S_{\alpha}^\e({\bf z})$.

\beq
\int_{S_\e(b_0 D_{\alpha})} \phi \circ S_\e^{-1}({\bf u})
\prod_{k=1}^N\frac{ u_k}{(S_\e^{-1} ({\bf u}))_k}
\det (S_{\e}^{-1})'({\bf u})
\prod_{k=1}^N\frac{1}{u_k -w_k}
du_1 \wedge \ldots \wedge du_N
\la{formintegration2}
\eeq

where $(S_{\e}^{-1})'$ is the complex derivative and so is holomorphic
in ${\bf u}$.
To apply Cauchy's formula we have to integrate over a product of cycles
(each lying in $\C$).
The map $t \mapsto S_t \stackrel{\mathrm{def}}{=}S_\alpha^t$
is a homotopy between $S_\e$ and the product map
$S_0$ and avoids singularities of the integrand in
(\ref{formintegration2}) since for $\e$ small enough the  set
$\{ S_t(b_0 D_{\alpha}) \mid 0\leq t \leq \e \}$ has positive
distance (uniformly in $\Lambda$) from the set of singularities
$\bigcup_{k=1}^N\{ u\in D_{\alpha}: u_k = w_k\}$.
$S_0(b_0 D_\alpha ) = S_{0,1}(\partial D_{\alpha_1})
\times \ldots \times
S_{0,N}(\partial D_{\alpha_N})$ is a product of cycles and hence a cycle. 
The differential n-form in (\ref{formintegration2}) is a
cocycle because its coefficient is holomorphic. So
we get by Stokes' theorem

\beq
= \int_{S_0 (b_0 D_{\alpha})} \phi \circ S_\e^{-1}({\bf u})
\prod_{k=1}^N\frac{ u_k}{(S_\e^{-1} ({\bf u}))_k}
\det (S_{\e}^{-1})'({\bf u})
\prod_{k=1}^N\frac{1}{u_k -w_k}
du_1 \wedge \ldots \wedge du_N
\eeq

and by Cauchy's formula

\beq
= \phi \circ S_\e^{-1}({\bf w})
\prod_{k=1}^N\frac{ w_k}{(S_\e^{-1} ({\bf w}))_k}\:
\frac{1}{\det (S_{\e}'(S_\e^{-1}({\bf w})))}
\eeq

So (\ref{formintegration1}) is equal to

\beq
\sum_{\alpha}
\int_{K} \frac{d{\bf w}}{(2\pi\imath)^N} \frac{1}{{\bf w}} 
\psi ({\bf w}) \phi\circ (S_{\alpha}^\e)^{-1}({\bf w})
\frac{1}{\det (S^\e) '((S_{\alpha}^\e)^{-1}({\bf w}))}
\prod_{k=1}^{N} \frac{w_k}{((S_{\alpha}^\e)^{-1}({\bf w}))_k}
\eeq

For each index $\alpha$, the map 
$S_{\alpha}^\e$ gives rise to a coordinate transformation
${\bf u}=(S_{\alpha}^\e)^{-1}({\bf w})$.

\beq
=\sum_{\alpha}\int_{K_{\alpha}} \frac{d{\bf u}}{(2\pi\imath)^N}
\frac{1}{{\bf u}}\psi\circ S_{\alpha}^\e ({\bf u}) \phi ({\bf u})
\eeq

As $\psi\circ F =0$ outside $\bigcup_{\alpha} K_{\alpha}$ and the 
$K_{\alpha}$ are mutually disjoint this equals

\ba
& = & \int_{(S^1)^N} \frac{d{\bf u}}{(2\pi\imath)^N}
\frac{1}{{\bf u}}\psi\circ S({\bf u}) \phi ({\bf u}) \\
& = & \int_{(S^1)^N} d\mu^{N} \psi\circ S \, \phi
\ea

as was to be  shown.
\qed


{\bf Proof of Lemma \ref{lemmamult} }Consistency follows from

\ba
\pi_{\Lambda_3} (g^1 \phi )_{\Lambda_4}
& = &
\pi_{\Lambda_3}\circ\pi_{\Lambda_4}
(g^1 \phi_{\Lambda \cup \Lambda_4} )\\
& = & 
\pi_{\Lambda_3}
(g^1 \phi_{\Lambda \cup \Lambda_4} )\nn\\
& = & 
\pi_{\Lambda_3}
(g^1 \phi_{\Lambda \cup \Lambda_3} )\nn\\
& = & 
(g^1 \phi )_{\Lambda_3}\nn
\ea

for all $\Lambda_3 \subset \Lambda_4 \in \mathcal{F}$.

As $g^1$ depends only on the $\Lambda_1$-coordinates we have

\ba
\|(g^1 \phi )_{\Lambda_1 \cup \Lambda} \|_{\Lambda_1 \cup \Lambda}
& = &
\|g^1 \phi_{\Lambda_1 \cup \Lambda} \|_{\Lambda_1 \cup \Lambda}\\
& \leq &
\| g^1 \|_{\Lambda_1}
\|\phi_{\Lambda_1 \cup \Lambda} \|_{\Lambda_1 \cup \Lambda}\nn\\
& \leq &
\| g^1 \|_{\Lambda_1}
\vth^{-|\Lambda_1| -|\Lambda|}
\|\phi\|_\vth\nn
\ea

and so

\beq
\vth^{|\Lambda|}
\| g^1 \phi )_{\Lambda} \|_{\Lambda}
\leq 
\| g^1\|_{\Lambda_1} \vth^{-|\Lambda_1|}\|\phi\|_\vth
\eeq

and
\beq
\| g \phi\|_\vth 
\leq 
\| g^1\|_{\Lambda_1} \vth^{-|\Lambda_1|}\|\phi\|_\vth
\eeq

For $\Lambda_1$ fixed the product is continuous in both factors.

(2.) follows from

\ba
((g^1 g^2)\phi )_{\Lambda}
& = &
\pi_{\Lambda}
(g^1_{\Lambda_1} g^2_{\Lambda_2} 
\phi_{\Lambda \cup \Lambda_1 \cup \Lambda_2})\\
& = &
\pi_{\Lambda}
(g^1_{\Lambda_1} 
\pi_{\Lambda \cup \Lambda_1}
(g^2_{\Lambda_2} 
\phi_{\Lambda \cup \Lambda_1 \cup \Lambda_2}))\nn\\
& = &
\pi_{\Lambda}
(g^1_{\Lambda_1} 
\pi_{\Lambda \cup \Lambda_1}
(g^2\phi))\nn\\
& = &
(g^1 (g^2\phi ))_{\Lambda}\nn
\ea

To see (3.) we note that the projection of the product of $g^1$
and $g^2$ is

\beq
\pi_{\Lambda}(g^1 g^2)
=
\pi_{\Lambda}(g^1_{\Lambda_1}  g^2_{\Lambda_2} )
\eeq

and the product in the sense of (\ref{defprod}) has
$\Lambda$-marginal

\ba
\pi_{\Lambda}(g^1 g^2)
& = &
\pi_{\Lambda}(g^1_{\Lambda_1} g^2_{\Lambda \cup \Lambda_2})\\
& = &
\pi_{\Lambda}(g^1_{\Lambda_1} g^2_{\Lambda_2})\nn
\ea

as $g^2$ does not depend on $\Lambda\setminus\Lambda_2$-coordinates.

If $\Lambda_1 \subseteq \Lambda_2$ then

\ba
g_{\Lambda_2} (g^1 \phi )_{\Lambda_2}
& = &
g_{\Lambda_2} g^1 \phi_{\Lambda_2}\\
& = &
(g^1 g)_{\Lambda_2} \phi_{\Lambda_2}\nn
\ea

and so (4.) follows from

\ba
(g^1 \phi ) (g)
& = &
\lim_{\Lambda_2 \to \Z^d}
\int_{(S^1)^{\Lambda_2}} d \mu^{\Lambda_2} g_{\Lambda_2}
(g^1 \phi )_{\Lambda_2}\\
& = &
\lim_{\Lambda_2 \to \Z^d}
\int_{(S^1)^{\Lambda_2}} d \mu^{\Lambda_2} (g^1 g)_{\Lambda_2}
\phi_{\Lambda_2}\nn\\
& = &
\phi (g^1 g)\nn
\ea

\ba
\| g^1 \phi\|_{var}
& = &
\lim_{\Lambda \to \Z^d}
\int_{(S^1)^{\Lambda}} d \mu^{\Lambda}
| (g^1 \phi )_{\Lambda}|\\
& = &
\lim_{\Lambda \to \Z^d \atop \Lambda \supset\Lambda_1}
\int_{(S^1)^{\Lambda}} d \mu^{\Lambda}
|g^1| |\phi_{\Lambda}|\nn\\
& \leq &
\| g^1\|_{\Lambda_1} \|\phi\|_{var}\nn
\ea
\qed


{\bf Proof of Proposition \ref{propsplitting} }
We get recursively

\ba
\lefteqn{
\frac{1}{f_p\circ T_{p,l}^{\e}({\bf z})-w_p}
\frac{f_p\circ T_{p,l}^{\e}({\bf z})}{z_p}}\\
& = &
\frac{1}{f_p\circ T_{p,l-1}^{\e}({\bf z})-w_p}
\frac{f_p\circ T_{p,l-1}^{\e}({\bf z})}{z_p} \nn\\
& & \,+
\frac{w_p}{z_p}
\frac{f_p\circ T_{p,l-1}^{\e}({\bf z}) -
f_p\circ T_{p,l}^{\e}({\bf z})}
{(f_p\circ T_{p,l-1}^{\e}({\bf z})-w_p)
(f_p\circ T_{p,l}^{\e}({\bf z})-w_p)} \nn\\
& = &
\frac{1}{f_p (z_p)-w_p}
\frac{f_p ({\bf z})}{z_p} 
+
\frac{w_p}{z_p}
\sum_{k=1}^{l}
\frac{f_p\circ T_{p,k-1}^{\e}({\bf z}) -
f_p\circ T_{p,k}^{\e}({\bf z})}
{(f_p\circ T_{p,k-1}^{\e}({\bf z})-w_p)
(f_p\circ T_{p,k}^{\e}({\bf z})-w_p)}
\nn
\ea

The estimate (\ref{formestimatebetak}) yields uniform convergence 
of this sum as $l\to\infty$.
So we get (\ref{formsplit}).
\qed

In (\ref{formestimatepi1lc}) we estimate the norm of the operator
corresponding to one particular configuration in terms of its different
kinds of lines and triangles. Now we have to bound sums over
all such configurations as they arise in expansions for the full operators.
For this we use our analysis and some combinatorics at the same time. 
The idea is that a configuration of a given length must have at least 
a certain number of triangles and r-chains that lead to small factors
in the estimates. In fact some special r-chains could not be replaced 
by h-chains in the configuration as we would get the zero operator.

\vspace{0.3cm}
{\bf Definition}
A maximal r-chain going from an apex downwards to the next base or
bottom point is called an \emph{a-r-chain}. (If the apex coincides
with a base or bottom point the a-r-chain has length zero.)

The \emph{a-r-length} of a configuration $\mathcal{C}$ is the sum of
the lenghts of all its a-r-chains plus the number of its triangles,
i.e.\ if $\mathcal{C}$ has   $|n_\beta |$ triangles with corresponding
a-r-chains of length $l_1, \ldots ,l_{|n_\beta |}$ then

\ba
\mbox{a-r-length} (\mathcal{C}) & \stackrel{\mathrm{def}}{=} & l_1+
\cdots +l_{|n_\beta |} + |n_\beta | \\
 & = & (l_1+1)+\cdots +(l_{|n_\beta |}+1)
\nn
\ea

(In particular $\mbox{a-r-length} (\mathcal{C}) \geq |n_\beta |$.)

We call a maximal r-chain going from a base point $(p,t)$ of a triangle
to $(p,-N)$ 
(such that $(p,-N)$ is not a base point of another triangle)
a \emph{u-r-chain} (upwards going r-chain),
a maximal r-chain going downwards from a basepoint a
\emph{d-r-chain} (\emph{d-h-chains} are defined analogously),
and a maximal r-chain going from
a bottom point $(p,0)$ to $(p,-N)$ an \emph{l-r-chain} (long r-chain).
The configuration in Figure \ref{figexampleconfiguration} has length 3, a-r-length 6,
only one a-r-chain of positive length from $(6,-2)$ to $(6,-1)$,
only one u-r-chain of positive length from $(3,-3)$ to $(3,-2)$,
and only one l-r-chain from $(1,-3)$ to $(1,0)$.

\vspace{0.3cm}
We prepare the proofs of Theorem \ref{theoremexistencepfo} and
Proposition \ref{propexpfo} in the following
technical proposition that provides the basic analysis and
combinatorics for all other proofs.


\begin{proposition}
For sufficiently small $\vth$, $\e$ and big $c_2$ and $N$ we
have for all 
$\Lambda_1 \subseteq \Lambda_2 \in \mathcal{F}$ 
the following bound for the terms in the expansion
of (\ref{formpipfo2})
for $\pi_{\Lambda_1} \circ\mathcal{L}^N_{F^{\Lambda_2}
\circ T^{\Lambda_2, \e}}$
with constants $c_{19}$, $c_{20}$:
\la{propetalaw}
\end{proposition}

\begin{enumerate}
\item
\beq
\sum_{\mathcal{C}: \mbox{\tiny length}(\mathcal{C})=N}
\left\| \pi_{\Lambda_1} \circ\mathcal{L_C} \right\|_{L\left(
\left( \mathcal{H}_{\Lambda_2 ,\vth},\|\cdot\|_{
\Lambda_2 ,\vth}\right),
\left( \mathcal{H}_{\Lambda_1 ,\vth},\|\cdot\|_{
\Lambda_1 ,\vth}\right)
\right)}
\leq c_{19} \tilde{\eta}^N
\eeq

\emph{with $\tilde{\eta}\stackrel{\mathrm{def}}{=}
\sqrt {\eta} < 1$}

\item
\beq
\left\| \pi_{\Lambda_1} \circ\mathcal{L}^N_{F^{\Lambda_2}
\circ T^{\Lambda_2 ,\e}}
\right\|
_{L\left(
\left( \mathcal{H}_{\Lambda_2 ,\vth},\|\cdot
\|_{\Lambda_2 ,\vth}\right),
\left( \mathcal{H}_{\Lambda_1 ,\vth},\|\cdot
\|_{\Lambda_1 ,\vth}\right)
\right)}
\leq c_{20}
\eeq
\end{enumerate}

{\bf Proof }
\begin{enumerate}
\item
We fix $0\leq K \leq |\Lambda_1 |$ and 
$\Lambda_3 \subseteq \Lambda_1$ with $|\Lambda_3 |=K$ (so there are 
${|\Lambda_1 | \choose K}$ possible choices for $\Lambda_3$)
and 
want to estimate the number of configurations $\mathcal{C}$ such that
$\Lambda_{\mathcal{C}} = \Lambda_3$.
So let us consider such a configuration.
We call the triangles whose apex lies at, or whose a-r-chain ends in,
$\Lambda_3 \times\{ 0\}$,
\emph{root triangles}.
We can assign to $\mathcal{C}$ a graph as follows: We start with 
a star graph with a star point labelled $(0)$ and $K$ leaves,
labelled $(0,1),\ldots ,(0,K)$. These leaves are in bijection with
$\Lambda_3 \times\{ 0\}$.
Now we add successively for each $l$-triangle in $\mathcal{C}$
a small \emph{l-tree} (a star graph with one star point and $v(l)$
leaves) to the graph and label the new vertices:
If an $l$-triangle lies with its apex or ends 
with its a-r-chain on a basepoint of another triangle 
(for that we have
already assigned a small tree) or on a point in $\Lambda_3 \times\{ 0\}$
(this point is labelled say $s=(s_1,\ldots ,s_n)$)
we add a small $l$-tree to the graph
by identifying its star point with 
$s$ and label the $v(l)$ new leaves in the graph
$(s_1,\ldots ,s_n,1)$,\ldots ,$(s_1,\ldots ,s_n,v(l))$.
Since, for example, an apex could coincide with more than one
other triangle's basepoint we introduce a linear order on the set
of tuples (and so on the set of vertices of the labelled graph):

We say $s=(s_1,\ldots ,s_n) \prec t=(t_1,\ldots ,t_m)$ if
$n<m$ and $s_i=t_i$ for $1 \leq i \leq n$ or if
$s_i=t_i$ ($1 \leq i \leq k)$ and $s_k < t_k$ for some $k$.

In our successive assignment of triangles to small trees we always choose
the next triangle such that the corresponding small tree is attached
 to the smallest (wrt.\ $\prec$) labelled leaf in the graph. This also
defines a unique choice of the triangle and the leaf where we 
attach the small tree.
So every 
$\mathcal{C}$ is completely determined by its corresponding labelled 
graph and the length of its a-r-chains. Note that it is
not the case that
for every graph together with a choice of lengths for the particular 
a-r-chains there was a corresponding configuration, but at least we have
found an injection between these two data.

For the configuration in Figure \ref{figexampleconfiguration},
for example, we get the following labelled graph:


\begin{figure}[h]
\begin{center}
\begin{minipage}[b]{\linewidth}
\centering 
\epsfig{file=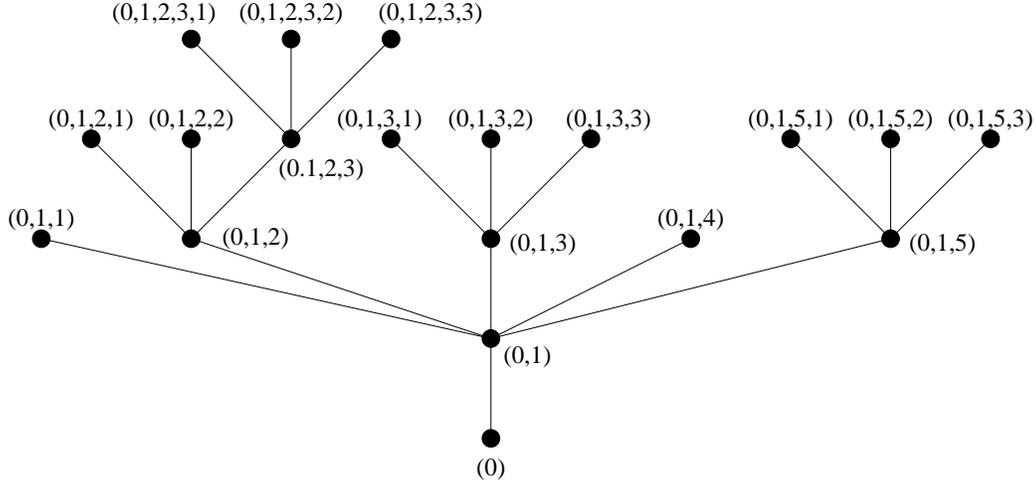,width=\linewidth}
\end{minipage}
\caption{\la{figlabelledgraph}The labelled graph for the configuration 
in Figure \ref{figexampleconfiguration}}
\end{center}
\end{figure}

If $n_{\beta ,k}$ is the total number of $k$-triangles, the number of such
corresponding sets of graphs is not greater than 
$4^K \prod_{k=1}^{\infty} c_{21}^{k^d n_{\beta ,k}}$ 
(by Lemma \ref{lemmatree}, see below).
As mentioned above we have for each of the $|n_\beta |$ a-r-chains a length
$0\leq l_i < \infty$.
The a-r-length is
\beq
L=(l_1+1)+\cdots +(l_{|n_\beta |}+1).
\la{formL}
\eeq

So $L\geq |n_\beta |$. 
For a given $n_\beta$ with $|n_\beta |\geq 1$ and $L\geq 1$ 
there are 
${L-1 \choose |n_\beta |-1}$ different choices of 
$(l_1,\ldots , l_{|n_\beta |} )$ that satisfy (\ref{formL}). 
For $|n_\beta |=0$ we have $L=0$
and the (unique) configuration without triangles or r-lines.
So in any case the number of choices is bounded from above by
${L \choose |n_\beta |}$.
The integration over these $|n_\beta |$
a-r-chains leads to a factor $c_r^{|n_\beta |} \eta^L$ in our estimates
(cf.\ (\ref{formestimatepi1lc})).

\item
There are choices between d-r-chains and d-h-chains 
in the configuration.

There are not more than $\sum_{k=1}^{\infty} (3k)^d n_{\beta ,k}$ 
base points for which
we can choose between a d-h-chain (giving factor $c_h$ in our estimates)
and a d-r-chain (giving factor at most $c_r \eta$). So the total sum
over these combinations is bounded from above by
\[(c_h + c_r \eta )^{\sum_{k=1}^{\infty} (3k)^d n_{\beta ,k}}
\leq
\prod_{k=1}^{\infty}
\left( \exp (c_{22} k^d)\right)^{n_{\beta ,k}}
\]

\item
There are choices between u-r-chains and u-h-chains 
in the configuration.

There are not more than $\sum_{k=1}^{\infty} (3k)^d n_{\beta ,k}$
basepoints.
To each of them we can attach either a u-h-chain, giving a factor
$c_h$, or a u-r-chain, giving a factor
$c_r \eta^{\max \{ 0, N-L \} }$,
because if $N-L >0$, such a u-r-chain cannot have length 
smaller than $N-L$, for otherwise it would not end in
$\Lambda_2 \times\{ -N\}$.
If $N-L > 0$ there must be at least one u-r-chain, so we get in total 
a factor not greater than
\beq
(c_h + c_r )^{\sum_{k=1}^{\infty} (3k)^d n_{\beta ,k}}
\eta^{\max \{ 0,N-L\}}
=
\prod_{k=1}^{\infty} \left(\exp (c_{23} k^d)\right)^{n_{\beta ,k}}
\eta^{\max \{ 0,N-L\}}
\eeq

\item
There are only choices left between l-h-chains and l-r-chains in
$(\Lambda_1\setminus \tilde{\Lambda}_{\mathcal{C}}) 
\times \{-N,\ldots ,0\}$,
giving factor $c_h$ or $c_r \eta^N$ respectively.
Let $l$ $(0\leq l\leq |\Lambda_1 \setminus \tilde{\Lambda}_{\mathcal{C}}|
\leq |\Lambda_1 |-K)$ denote the number of l-r-chains in such a
choice.
For given $l$ there are
${|\Lambda_1 \setminus \tilde{\Lambda}_{\mathcal{C}}| \choose l }
\leq {|\Lambda_1 | -K \choose l }$
different subsets $\Lambda_r$ of $\Lambda_1 \setminus 
\tilde{\Lambda}_{\mathcal{C}}$
of cardinality $l$ (that corresponds to a particular choice of exactly
$l$  l-r-chains.)
The configuration $\mathcal{C}$  is determined by all the choices 
mentioned up to now.

\end {enumerate}

In the configuration $\mathcal{C}$ there are h-chains at 
points with $\Z^d$-coordinate in\\
$\Lambda_1 \setminus (\tilde{\Lambda}_{\mathcal{C}}
\cup \Lambda_r )$. The operator
$\mathcal{L_C}$ acts on $\phi_{\Lambda_2}$ by integration
over these coordinates. 
So for the uniform estimate of
$\mathcal{L_C} \phi_{\tilde{\Lambda}}$ we will use
(\ref{formestimatephilambdaclambdar}).

Therefore we have the estimate

\ba
\lefteqn{
\vth^{|\Lambda_1 |}
\sum_{\mathcal{C}: \mbox{\tiny length}(\mathcal{C})=N}
\left\| \pi_{\Lambda_1} \circ\mathcal{L_C} \phi_{\Lambda_2}
\right\|_{\Lambda_1}}
\la{formestimate0}\\
& \leq &
\vth^{|\Lambda_1 |}
\sum_{K=0}^{|\Lambda_1 |} {|\Lambda_1 | \choose K}
\sum_{n_\beta \atop K\leq |n_\beta|<\infty} 4^K
\prod_{k=1}^{\infty} \left(\exp (c_{21} k^d)\right)^{n_{\beta ,k}}
(c_1 \e)^{|n_\beta|} 
\la{formestimate2}\\
&&\times
\prod_{k=1}^{\infty} \left(\exp (-c_2 k^d)\right)^{n_{\beta ,k}}
\sum_{L=|n_\beta |}^\infty {L \choose |n_\beta |}
c_r^{|n_\beta |} \eta^L
\prod_{k=1}^{\infty} \left(\exp (c_{22} k^d)\right)^{n_{\beta ,k}}
\nn\\ 
&&\,\times
\prod_{k=1}^{\infty} \left(\exp (c_{23} k^d)\right)^{n_{\beta ,k}}
\eta^{\max \{ 0,N-L\}}
\sum_{l=0}^{|\Lambda_1 | -K} {|\Lambda_1 | -K \choose l}
(c_r\eta^N)^l
\nn\\ 
&& \,\times
 c_h^{|\Lambda_1 | -K-l} 
\vth^{-l} \prod_{k=1}^{\infty} \vth^{-(3k)^d n_{\beta ,k}}
\|\phi\|_{\Lambda_2 ,\vth} 
\nn\\
& \leq &
\vth^{|\Lambda_1 |}
\sum_{K=0}^{|\Lambda_1 |} {|\Lambda_1 | \choose K}
\sum_{n_\beta \atop K\leq |n_\beta|<\infty} 4^K
(c_1 \e c_r)^{|n_\beta|}
\la{formestimate}
\\ \,
&&\times
\prod_{k=1}^{\infty}
\exp ( (c_{21} -c_2 +c_{22}+c_{23} - 3^d \ln \vth ) k^d)^{n_{\beta ,k}}
\nn\\ \,
&&\times
\sum_{L=|n_\beta |}^\infty {L \choose |n_\beta |}
\eta^{\max \{ N ,L\}}
(\vth^{-1} c_r \eta^N + c_h)^{|\Lambda_1 |-K}
\|\phi\|_{\Lambda_2 ,\vth}.
\nn
\ea

We set $\e_1 \stackrel{\mathrm{def}}{=} 4 \e c_1 c_r$
and $\e_2 \stackrel{\mathrm{def}}{=} \sqrt{\e_1}$.
Then we have 
$\e_1^{|n_\beta|} \leq \e_2^K \e_2^{|n_\beta|}$.
We set 
$\tilde{c_2}\stackrel{\mathrm{def}}{=}
c_2 -c_{21} -c_{22}-c_{23} +3^d \ln \vth$.
Then $\tilde{c_2} > 0$ if 
\beq
c_2 > c_{21} +c_{22}+c_{23}- 3^d \ln \vth 
\la{formconditionc2}
\eeq

(We assume this condition on the decay of the coupling.)
Further we split
$\eta^{\max \{ N ,L\}} \leq \tilde{\eta}^L \tilde{\eta}^N$
with $\tilde{\eta}=\sqrt{\eta}$.
Then (\ref{formestimate}) can be bounded as follows:

\ba
\lefteqn{}& \leq &
\sum_{K=0}^{|\Lambda_1 |} {|\Lambda_1 | \choose K}
(c_r \eta^N + \vth c_h)^{|\Lambda_1 |-K} \e_2^K
\sum_{n_\beta \atop K\leq |n_\beta|<\infty}
\sum_{L=|n_\beta |}^\infty {L \choose |n_\beta |}
\tilde{\eta}^L \e_2^{|n_\beta|} \\ 
&& \,\times
\prod_{k=1}^{\infty}
\left( \exp ( -\tilde{c_2} k^d)\right)^{n_{\beta ,k}}
\|\phi\|_{\Lambda_2 ,\vth} \tilde{\eta}^N
\nn\\
& \leq &
(c_r \eta^N + \vth c_h + \e_2)^{|\Lambda_1 |}
\sum_{L=0}^\infty \sum_{n=0}^L {L \choose n} {\tilde{\eta}}^L \e_2^n
\sum_{n_\beta \atop |n_\beta|=n}
\prod_{k=1}^{\infty}
\left( \exp ( -\tilde{c_2} k^d)\right)^{n_{\beta ,k}}
\nn\\
&&\,\times
\|\phi\|_{\Lambda_2 ,\vth} \tilde{\eta}^N
\nn
\ea

We have 
\beq
\sum_{n_\beta \atop |n_\beta|=n}
\prod_{k=1}^{\infty}
\left( \exp ( -\tilde{c_2} k^d)\right)^{n_{\beta ,k}}
\leq
\prod_{k=1}^{\infty}
\sum_{n_{\beta ,k}=0}^\infty
\left( \exp ( -\tilde{c_2} k^d)\right)^{n_{\beta ,k}}
\eeq

and the last infinite product converges
(to $c_{24}$ say) since for $k$ sufficiently large
$\exp ( -\tilde{c_2} k^d) < \frac{1}{2}$ and
$\sum_{n_{\beta ,k}=0}^{\infty}
\left( \exp ( -\tilde{c_2} k^d)\right)^{n_{\beta ,k}}
\leq 1 + 2 \exp ( -\tilde{c_2} k^d)$
and 
$\sum_{k=0}^{\infty} \exp ( -\tilde{c_2} k^d) < \infty$ 
(Recall 
$\prod_{k=1}^{\infty} (1+u_k)$ \emph{convergent}
$\Leftarrow
\sum_{k=1}^{\infty} |u_k | < \infty$.)

\ba
\lefteqn{ }
& \leq &
(\e_2 + c_r \tilde{\eta}^N + c_h \vth)^{|\Lambda_1 |} c_{24}
\sum_{L=0}^{\infty} (\e_2 +\tilde{\eta})^L
\|\phi\|_{\Lambda_2 ,\vth} \tilde{\eta}^N
\nn \\
& \leq &
(\e_2 + c_r \tilde{\eta}^N + c_h \vth)^{|\Lambda_1 |}
\frac{1}{1-\e_2-\tilde{\eta}} c_{24}
\|\phi\|_{\Lambda_2 ,\vth} \tilde{\eta}^N
\la{formestimatecretaN}\\
& \leq &
c_{19} \tilde{\eta}^N \|\phi\|_{\Lambda_2 ,\vth}
\la{formestimate9}
\ea

for $\vth$ and $\e$ sufficiently small and $N$ sufficiently large.
This also holds for $\Lambda \subset \Lambda_1$. So (1.) is proved.

If $\mathcal{C}$ is a non-zero configuration of length $0\leq m< N$
in the expansion of $\pi_{\Lambda_1} \circ
\mathcal{L}^N_{F^{\Lambda_2}\circ
T^{\Lambda_2 ,\e}}$ it has no l-r-chains. So this time we have 
$l(\mathcal{C})=0$.
Using the splitting
$\eta^L \leq \tilde{\eta}^L \tilde{\eta}^m$
we get in a similar way

\ba
\lefteqn{\vth^{|\Lambda_1 |}
\sum_{\mathcal{C}: \mbox{\tiny length}(\mathcal{C})=m, \atop
l(\mathcal{C})=0}
\left\| \pi_{\Lambda_1} \circ\mathcal{L_C} \phi_{\Lambda_2} 
\right\|_{\Lambda_1}}
\la{formestimatel0}\\
& \leq &
\vth^{|\Lambda_1 |}
\sum_{K=0}^{|\Lambda_1 |} {|\Lambda_1 | \choose K}
\sum_{n_\beta \atop K\leq |n_\beta|<\infty} 4^K
\prod_{k=1}^{\infty} \left(\exp (c_{21} k^d)\right)^{n_{\beta ,k}}
\nn\\
&&\,\times
(c_1 \e)^{|n_\beta|}
\prod_{k=1}^{\infty} \left(\exp (-c_2 k^d)\right)^{n_{\beta ,k}}
\sum_{L=|n_\beta |}^\infty {L \choose |n_\beta |}
c_r^{|n_\beta |} \tilde{\eta}^L
\prod_{k=1}^{\infty} \left(\exp (c_{22} k^d)\right)^{n_{\beta ,k}}
\nn\\
&&\,\times
\prod_{k=1}^{\infty} \left(\exp (c_{23} k^d)\right)^{n_{\beta ,k}}
c_h^{|\Lambda_1 | -K} 
\prod_{k=1}^{\infty} \vth^{-(3k)^d n_{\beta ,k}}
\|\phi\|_{\Lambda_2 ,\vth} \tilde{\eta}^N
\nn\\
& \leq &
\sum_{K=0}^{|\Lambda_1 |} {|\Lambda_1 | \choose K}
(c_h \vth)^{|\Lambda_1 | -K}
\sum_{n_\beta \atop K\leq |n_\beta|<\infty} 
(c_1 \e 4 c_r)^{|n_\beta|}
\prod_{k=1}^{\infty} \left(\exp (- \tilde{c_2} k^d)\right)^{n_{\beta ,k}}
\nn\\
&&\,\times
\sum_{L=|n_\beta |}^\infty {L \choose |n_\beta |}\tilde{\eta}^L
\tilde{\eta}^m \|\phi\|_{\Lambda_2 ,\vth}
\nn\\
& \leq &
\sum_{K=0}^{|\Lambda_1 |} {|\Lambda_1 | \choose K}
(c_h \vth)^{|\Lambda_1 | -K} \e_2^K
\sum_{n_\beta \atop K\leq |n_\beta|<\infty}
\sum_{L=|n_\beta |}^\infty {L \choose |n_\beta |}\tilde{\eta}^L
\e_2^{|n_\beta |}
\nn\\
&&\,\times
\prod_{k=1}^{\infty} \left(\exp (- \tilde{c_2} k^d)\right)^{n_{\beta ,k}}
\tilde{\eta}^m \|\phi\|_{\Lambda_2 ,\vth}
\nn\\
& \leq &
(\e_2 + c_h \vth)^{|\Lambda_1 |}
\frac{1}{1-\e_2-\tilde{\eta}} c_{25} \tilde{\eta}^m
\|\phi\|_{\Lambda_2 ,\vth} 
\nn \\
& \leq &
c_{26} \tilde{\eta}^m \|\phi\|_{\Lambda_2 ,\vth}
\ea

Again this also holds for $\Lambda\subset\Lambda_1$ and so 

\beq
\vth^{|\Lambda_1 |}
\sum_{\mathcal{C}: \mbox{\tiny length}(\mathcal{C})=m, \atop
l(\mathcal{C})=0}
\left\| \pi_{\Lambda_1} \circ\mathcal{L_C} \phi_{\Lambda_2} 
\right\|_{\Lambda_1 ,\vth}
\leq
c_{26} \|\phi\|_{\Lambda_2 ,\vth} \tilde{\eta}^m
\la{formestimatewithc26}
\eeq

Therefore

\ba
\left\| \pi_{\Lambda_1} \circ\mathcal{L}^N_{F^{\Lambda_2}
\circ T^{\Lambda_2 ,\e}} 
\right\|_{L((\mathcal{H}_{\Lambda_2 ,\vth} , 
\|\cdot\|_{\Lambda_2 ,\vth} ),
(\mathcal{H}_{\Lambda_1 ,\vth }
\|\cdot\|_{\Lambda_1 ,\vth} )}
& \leq & 
\sum_{m=0}^N c_{26} \tilde{\eta}^m \\
& \leq & 
\sum_{m=0}^\infty c_{26} \tilde{\eta}^m \nn\\
& \leq &
c_{20} \nn
\ea
\qed


\begin{lemma}
\begin{enumerate}
\item
The number of labelled tree graphs with exactly n edges is smaller
than $2^{2n}$
\item
The number of labelled tree graphs corresponding to configurations that
have exactly $n_{\beta ,k}$ $k$-triangles ($|n_{\beta ,k}| < \infty$) and
end (at time $0$) in \\ $\Lambda_3\times\{ 0\}$ and not in any smaller set
is bounded from above by\\
$4^{|\Lambda_3|} \prod_{k=1}^\infty c_{21}^{k^d n_{\beta ,k}}$
with $c_{21}=4^{3^d}$.
\end{enumerate}
\la{lemmatree}
\end{lemma}

{\bf Proof }
1.) For every labelled tree graph in question we can define a path
starting and ending at the root point $(0)$ and running through each edge
exactly twice in the following way.
>From a (labelled) vertex $t=(t_1,\ldots ,t_n)$ we go 
to the to the next greater (wrt.\ $\prec$) vertex where we haven't yet been
(\emph{going up}), or if this is not possible (i.e.\ $t$ is a leaf
or we have already been at all vertices 
$(t_1,\ldots ,t_{n+1})$) back to 
$(t_1,\ldots ,t_{n-1})$ (\emph{going down}).
So we return to $(0)$ after $2n$
steps. We encode the path in a word 
$(a_1,\ldots ,a_{2n})$ with
$a_i=1$ if we go up in the $i$th step and $a_i=0$ otherwise.
Obviously the labelled graph is uniquely determined by its word.
(Note that not every word of length $2n$ with symbols
"$0$" and "$1$" corresponds to such a labelled graph. 
But the map between these two data is injective.)
As there are 
$2^n$ words of length
$2n$ with at most two different symbols this  is also
an upper bound for the number of graphs in question.

To see (2.) we note that the number of edges in such a tree graph
is not greater than $K+\sum_{k=1}^\infty (3k)^d n_{\beta ,k}$.
\qed


{\bf Proof of Theorem \ref{theoremexistencepfo}}
The difference between 
$\pi_{\Lambda_1} \circ \mathcal{L}^N_{F^{\Lambda_2} 
\circ T^{\Lambda_2 ,\e}} \circ \pi_{\Lambda_2}$ and 
$\pi_{\Lambda_1} \circ \mathcal{L}^N_{F^{\Lambda_3} 
\circ T^{\Lambda_3 ,\e}} \circ \pi_{\Lambda_3}$ 
for $\Lambda_1 \subseteq \Lambda_2 \subseteq \Lambda_3 \in \mathcal{F}$
is due to the summands involving configurations that do not lie 
completely (with all its triangles) in 
$\Lambda_2 \times \{ 0, -1, \ldots \}$.
For those we have the lower bound for the spatial extension of the set of 
triangles:

\ba
b(\mathcal{C}) & \stackrel{\mathrm{def}}{=} &
\sum_{k=1}^{\infty} k n_{\beta ,k}
\la{formspacialextension}\\
& \geq &
\mbox{dist}(\Lambda_1 ,\bL)
\nn
\ea

As the analysis in the proof of Proposition \ref{propetalaw} 
shows we have in the  estimate for each such configuration a factor

\ba
\lefteqn{
\prod_{k=1}^{\infty} \left( \exp (-\tilde{c_2} k^d)
\right)^{n_{\beta ,k}}}\\
& \leq &
\prod_{k=1}^{\infty} \left[ \exp (-(\tilde{c_2}-\xi ) k^d)
\right]^{n_{\beta ,k}}
\prod_{k=1}^{\infty} \left( \exp (- \xi k n_{\beta ,k})\right) \nn\\
& \leq & 
\prod_{k=1}^{\infty} \left[ \exp (-(\tilde{c_2}-\xi ) k^d)
\right]^{n_{\beta ,k}}
\exp \left(- \xi \mbox{dist}(\Lambda_1 ,\bL)\right)
\nn
\ea

If we take $\xi >0$ small enough
we can take out a factor
$\exp \left(- \xi \mbox{dist}(\Lambda_1 ,\bL)\right)$
and do the analysis with the remaining factor as before 
since $\tilde{c_2}-\xi >0$.
So we get

\ba
\|\pi_{\Lambda_1} \circ \mathcal{L}^N_{F^{\Lambda_2} 
\circ T^{\Lambda_2 ,\e}} \circ \pi_{\Lambda_2}
-
\pi_{\Lambda_1} \circ \mathcal{L}^N_{F^{\Lambda_3} 
\circ T^{\Lambda_3 ,\e}} \circ \pi_{\Lambda_3}
\|_{L\left(\left(\mathcal{H}_{\vth} ,
\|\cdot\|_{\vth}\right),
\left( \mathcal{H}_{\Lambda_1 ,\vth} ,
\|\cdot\|_{\Lambda_1 ,\vth}\right)\right)}  \nn\\
\leq 
c_{27} \exp \left(- \xi \mbox{dist}(\Lambda_1 ,\bL)\right)
\ea

with some constant $c_{27}$ and the limit in (\ref{formpi1pfo}) exists.
The second statement in (1.) follows from
(\ref{formestimate0}) and (\ref{formestimatel0})
with $\vth$ replaced by a sufficiently small $\tvth$.
For example,
(\ref{formestimate0}) becomes

\ba
\lefteqn{
\tilde{\vth}^{|\Lambda_1 |}
\sum_{\mathcal{C}: \mbox{\tiny length}(\mathcal{C})=N}
\left\| \pi_{\Lambda_1} \circ\mathcal{L_C} \phi_{\Lambda_2}
\right\|_{\Lambda_1}}\\
& \leq &
c_{28} (\e_2 + c_r \tilde{\eta}^n \frac{\tilde{\vth}}{\vth}
+c_h \tilde{\vth})^{|\Lambda_1 |}
\|\phi\|_{\Lambda_2 ,\vth} \tilde{\eta}^N \nn
\ea

and the term in brackets is smaller than $1$ if $\tilde{\vth}$
and $\frac{\tilde{\vth}}{\vth}$ are small enough.
The statement
for systems with finite-range interaction follows from the fact that
in that case all limits are already attained for some sufficiently large
$\Lambda_2\in\mathcal{F}$ and that all considered sums are finite.

(2.) follows from (3.) and (5.) of Proposition \ref{propexpfo}.
\qed


{\bf Proof of Proposition \ref{propexpfo}}
With the same argument as in the proof of (1.) in Theorem
\ref{theoremexistencepfo}
we see that the right-hand side term in
(\ref{formpi1pfoN}) differs from the operator in (\ref{formpipfo2}) 
only in summands
for $\mathcal{C}$ with
$b(\mathcal{C})
\geq
\mbox{dist}(\Lambda_1 ,\bL)$. So the difference is bounded by
$c_{29} \exp \left(- \xi \mbox{dist}(\Lambda_1 ,\bL)\right)$ for
some $c_{29}>0$.

In order to prove (2.) we first observe that configurations 
$\mathcal{C}\in E_N (\Lambda_1)$ of length $\leq N-1$ 
extend canonically to $\mathcal{C'}\in E_{N+1} (\Lambda_1)$ 
with $\mathcal{L_C} = \mathcal{L_{C'}}$ 
because there are only h-lines
in the step from time $-N$ to $-N+1$. So we can extend $\mathcal{C}$
to $\mathcal{C'}$ on $\Lambda_2\times\{ -N-1,\ldots , 0\}$ 
(where $\Lambda_2$ is so big that $\Lambda_2\times\{ -N-1,\ldots , 0\}$
contains all triangles of $\mathcal{C}$) by
adding h-lines from $(p,-N-1)$ to $(p,-N)$ for all 
$p\in \Lambda_2$ and obviously 
$\mathcal{L_C} = \mathcal{L_{C'}}$.

Note that a configuration $\mathcal{C'}$ in 
$\Lambda_2\times\{ -N-1, 0\}$ of length $\leq N-1$ is the extension
in the above sense of a (uniquely defined) $\mathcal{C}$.

So in the difference (\ref{formcauchycritpfo}),
all terms $\mathcal{L_C}$ with
length $\leq N-1$ are cancelled. Using (1.) of Proposition \ref{propetalaw},
(\ref{formestimatewithc26})
and (1.) of this proposition we get
for all $\Lambda_1 \in\mathcal{F}$

\ba
\left\|\left(\pi_{\Lambda_1} \circ \mathcal{L}^N_{F \circ T^\e}
-\pi_{\Lambda_1} \circ \mathcal{L}^{N+1}_{F \circ T^\e}\right)
\phi\right\|_{\Lambda_1,\vth}
\la{formpetalaw}
& \leq & 
\left( c_{19} \tilde{\eta}^N + c_{20} \tilde{\eta}^N 
+ c_{19} \tilde{\eta}^{N+1}
\right) \|\phi\|_{\vth}
\nn\\
& \leq & 
c_{30} \tilde{\eta}^N \|\phi\|_{\vth}
\la{formestimatec30}
\ea

with $c_{30}$ independent of $\Lambda_1$. 
This proves (2.)

Recall that by Theorem \ref{theoremexistencepfo} the operators 
$\mathcal{L}^N_{F \circ T^\e} \in
L\left(\mathcal{H}_{\vth} ,\|\cdot
\|_{\vth}\right)$ are well defined for $N\geq N_0$
and, by part (2.), give rise to a Cauchy sequence.
With the same argument we see that the infinite sum in the 
definition of $\nu_\Lambda$ (cf.\ (\ref{defnu})) converges
and $\nu\in\mathcal{H}_\vth$. $\nu \geq 0$ and so
$\nu\in\mathcal{H}^{bv}$ will follow from (7.).

The difference in (\ref{formestimatepfoNminusmu}) is only due to
configurations of length $\geq N$ and can therefore be estimated
(as before in (\ref{formestimatec30})) by
$c \tilde{\eta}^N$, which proves (\ref{formestimatepfoNminusmu}).

For $\Lambda_1\in\mathcal{F}$,

\ba
\lefteqn{
\pi_{\Lambda_1} \circ \mathcal{L}^{N_2}_{F \circ T^\e} 
\circ \mathcal{L}^{N_1}_{F \circ T^\e} \phi} \\
& = &
\sum_{\mathcal{C}_2\in E_{N_2}(\Lambda_1)}
\pi_{\Lambda_1} \circ \mathcal{L}_{\mathcal{C}_2}
\left(\mathcal{L}^{N_1}_{F \circ T^\e} \phi \right) \nn\\
& = &
\sum_{\mathcal{C}_2\in E_{N_2}(\Lambda_1)}
\left(\pi_{\Lambda_1} \circ \mathcal{L}_{\mathcal{C}_2}
\circ \sum_{\mathcal{C}_1\in E_{N_2}(\Lambda(\mathcal{C}_2))}
\pi_{\Lambda(\mathcal{C}_2)}\circ\mathcal{L}_{\mathcal{C}_1}
\phi_{\Lambda(\mathcal{C}_1)}\right) \nn\\
& = &
\sum_{\mathcal{C}_2\in E_{N_2}(\Lambda_1) \atop
\mathcal{C}_1\in E_{N_2}(\Lambda(\mathcal{C}_2))}
\pi_{\Lambda_1} \circ \mathcal{L}_{\mathcal{C}_2\circ \mathcal{C}_1}
\phi_{\Lambda(\mathcal{C}_1)} \nn\\
& = &
\sum_{\mathcal{C}_3\in E_{N_1+N_2}(\Lambda_1)}
\pi_{\Lambda_1} \circ \mathcal{L}_{\mathcal{C}_3}
\phi_{\Lambda(\mathcal{C}_3)} \nn\\
& = &
\pi_{\Lambda_1} \circ 
\mathcal{L}^{N_1+N_2}_{F \circ T^\e} \phi . \nn
\ea

Note that we sum over infinitely many $\mathcal{C}_1 , \mathcal{C}_2$.
A priori, the distribution is only valid for finite partial sums. In terms
of configurations we `put $\mathcal{C}_1$ on $\mathcal{C}_2$'
to get $\mathcal{C}_3=\mathcal{C}_2\circ \mathcal{C}_1$ 
(which might be a zero configuration) and in fact 
such a splitting exists and is unique for every non-zero $\mathcal{C}_3$.
So the net of finite partial sums over $\mathcal{C}_3$ we get converges
to the infinite expansion (\ref{formpi1pfoN}) of the right-hand side
of (\ref{formsemigroup}) and (4.) is proved.
We have by (3.)
$\lim_{N \rightarrow \infty} \mathcal{L}^N_{F \circ T^\e} h
=\mu (h) = \nu$ and so by (4.)
$\mathcal{L}_{F \circ T^\e} \nu =\nu$ and also $\mu (\nu )=1$,
by (6.) which we will show below.
For any
$\phi\in \mathcal{H}_\vth$ with 
$\mathcal{L}_{F \circ T^\e} 
\phi = \phi$ and $\mu (\phi )=1$ we have

\beq
\phi = \lim_{N \rightarrow \infty} \mathcal{L}^N_{F \circ T^\e} \phi
=\mu (\phi ) \nu = \nu
\eeq

so (5.) is proved.

To prove (6.), we consider first the special case
$g \in \mathcal{C} ((S^1)^\Lambda )$.

\ba
\int_M d\mu \, g \circ S \, \phi
& = &
\lim_{\Lambda_1 \to \Z^d} \int_M d\mu \, g \circ S_{\Lambda_1} \phi 
\la{formadjoint}\\
& = &
\lim_{\Lambda_1 \to \Z^d} \int_{(S^1)^{\Lambda_1}}
d\mu^{\Lambda_1} g \circ S_{\Lambda_1} \phi_{\Lambda_1} \nn\\
& = &
\lim_{\Lambda_1 \to \Z^d} \int_{(S^1)^{\Lambda_1}}
d\mu^{\Lambda_1} g \,
\mathcal{L}_{F^{\Lambda_1} \circ T^{\Lambda_1 ,\e}}\phi_{\Lambda_1} \nn\\
& = &
\lim_{\Lambda_1 \to \Z^d} 
\int_{(S^1)^{\Lambda}}
d\mu^{\Lambda} g \,
\pi_\Lambda \circ 
\mathcal{L}_{F^{\Lambda_1} \circ T^{\Lambda_1 ,\e}} \pi_{\Lambda_1}
\phi \nn\\
& = &
\int_M d\mu \, g \mathcal{L}_{F \circ T^\e}\phi \nn
\ea

So (6.) is true for $g \in \mathcal{C} ((S^1)^\Lambda )$.
By assumption $\phi$ and so 
$\mathcal{L}_{F\circ T^\e} \phi$ are in $\mathcal{H}^{bv}$
(by part (8.) for whose proof we just use the above special case of (6.)),
i.e.\ they correspond to continuous linear
functionals. For any $g\in \mathcal{C} (M)$ the net
$(g_\Lambda )_{\Lambda\in\mathcal{F}}$ converges uniformly to $g$
as $\Lambda \to \Z^d$, as does 
$(g_\Lambda \circ S)_{\Lambda\in\mathcal{F}}$
to $g\circ S$. So by continuity of the integral operators
the equality also holds for $g$.
The special case (6.) follows from taking $g \equiv 1$.
For finite-range interaction the limits in (\ref{formadjoint}) are already
attained for sufficiently large $\Lambda_1\in\mathcal{F}$ and all 
the computations work with $\phi\in\mathcal{H}^{bv}$.

We have, by definition,
$(\mathcal{L}_{F \circ T^\e} \phi)_\Lambda
\stackrel{\mathrm{def}}{=}\lim_{\Lambda_1 \to \Z^d}
\pi_{\Lambda} \circ \mathcal{L}_{F^{\Lambda_1} 
\circ T^{\Lambda_1 ,\e}} \phi_{\Lambda_1}$.
If that was negative somewhere there would be a $\Lambda_1\in\mathcal{F}$
with $\pi_{\Lambda} \circ \mathcal{L}_{F^{\Lambda_1} 
\circ T^{\Lambda_1 ,\e}} \phi_{\Lambda_1}$ having negative values and we
could find a non-negative $g\in \mathcal{C} ((S^1)^\Lambda )$
such that

\beq
\int_{(S^1)^\Lambda } d \mu^\Lambda 
g \, \pi_\Lambda \circ
\mathcal{L}_{F^{\Lambda_1} 
\circ T^{\Lambda_1 ,\e}} \phi_{\Lambda_1}
< 0
\eeq

But by (6.) the integral equals

\beq
\int_{(S^1)^{\Lambda_1}} d \mu^{\Lambda_1}
g \circ S \, \phi_{\Lambda_1}
\geq
0
\eeq

So $\mathcal{L}_{F\circ T^\e}$ is non-negative. Finally (8.) follows from

\ba
\|\mathcal{L}_{F\circ T^\e} \phi\|_{var}
& = &
\sup_{\Lambda\in \mathcal{F}}
\sup_{g\in\mathcal{C}((S^1)^\Lambda ) \atop
\| g\|_\infty \leq 1}
\int_M d \mu \, g \, \mathcal{L}_{F\circ T^\e} \phi\\
& = &
\sup_{\Lambda\in \mathcal{F}}
\sup_{g\in\mathcal{C}((S^1)^\Lambda ) \atop
\| g\|_\infty \leq 1}
\int_M d \mu \, g \circ S \,\phi \nn\\
& \leq &
\sup_{\Lambda\in \mathcal{F}}
\sup_{g\in\mathcal{C}((S^1)^\Lambda ) \atop
\| g\|_\infty \leq 1}
\| g\|_\infty \| \phi\|_{var} \nn\\
& = &
\| \phi\|_{var}
\nn
\ea
\qed


{\bf Proof of Theorem \ref{propositiondecay1}}
\ba
\nu_{\Lambda_1 \cup \Lambda_2}
& = & \sum_{\mathcal{C}\in E(\Lambda_1 \cup \Lambda_2)}
\pi_{\Lambda_1 \cup \Lambda_2}\circ \mathcal{L_C} h
\la{formnu12}\\
& = &
\sum_{\mathcal{C} = \mathcal{C}_1 \cup \mathcal{C}_2 \atop
b(\mathcal{C})\leq 
\frac{1}{2}\dist}
(\pi_{\Lambda_1} \circ \mathcal{L}_{\mathcal{C}_1}h)
(\pi_{\Lambda_2} \circ \mathcal{L}_{\mathcal{C}_2}h) 
\nn\\\,
&& +
\sum_{\mathcal{C} \atop 
b(\mathcal{C}) >  \frac{1}{2}\dist}
\pi_{\Lambda_1 \cup \Lambda_2}\circ \mathcal{L_C} h
\nn
\ea 

In estimating the second summand we note that if we sum in formula
(\ref{formestimate0}) and (\ref{formestimatel0}) just over
$\mathcal{C}$ with
$b(\mathcal{C})\geq \frac{1}{2} \bdist$
we can take out from\\
$\prod_{k=1}^{\infty}\left( \exp (- \tilde{c_2} k^d)\right)^{n_{\beta ,k}}$
a factor $\exp (-\xi 
\frac{1}{2}) \bdist$
(like in the proof of Proposition \ref{propexpfo}). The rest 
of the analysis is as in the proof of Proposition \ref{propetalaw}. 
We can take $\xi$ such that $\exp (-\xi \frac{1}{2})=\kappa$
if $c_2$ is sufficiently large and get

\ba
\lefteqn{
\|
\sum_{\mathcal{C} \atop
b(\mathcal{C}) > \frac{1}{2}\dist}
\pi_{\Lambda_1 \cup \Lambda_2}\circ \mathcal{L_C} h
\|_{\Lambda_1 \cup \Lambda_2}}\\
& \leq & 
\kappa^{\dist} 
c_{31} \| h\|_\vth 
\vth^{- | \Lambda_1 | - | \Lambda_2 |}\\
& \leq & 
c_{32} \vth^{- | \Lambda_1 | - | \Lambda_2 |}
\kappa^{\dist}
\nn
\la{formestimate5}
\ea

We write for the first summand in (\ref{formnu12})

\ba
\lefteqn{
\sum_{\mathcal{C} = \mathcal{C}_1 \cup \mathcal{C}_2 \atop
b(\mathcal{C})\leq \frac{1}{2}\dist}
(\pi_{\Lambda_1} \circ \mathcal{L}_{\mathcal{C}_1}h)
(\pi_{\Lambda_2} \circ \mathcal{L}_{\mathcal{C}_2}h)}\\
\vspace{0.3cm}
& &
=
\nu_{\Lambda_1}\nu_{\Lambda_2}
-
\sum_{\mathcal{C} = \mathcal{C}_1 \cup \mathcal{C}_2 \atop
b(\mathcal{C}) > \frac{1}{2}\dist}
(\pi_{\Lambda_1} \circ \mathcal{L}_{\mathcal{C}_1}h)
(\pi_{\Lambda_2} \circ \mathcal{L}_{\mathcal{C}_2}h)
\nn
\ea

and estimate in a similar way

\beq
\|\sum_{\mathcal{C} = \mathcal{C}_1 \cup \mathcal{C}_2 \atop
b(\mathcal{C}) > \frac{1}{2}\dist}
(\pi_{\Lambda_1} \circ \mathcal{L}_{\mathcal{C}_1}h)
(\pi_{\Lambda_2} \circ \mathcal{L}_{\mathcal{C}_2}h)
\|_{\Lambda_1\cup\Lambda_2}
\leq
c_{33} \vth^{- | \Lambda_1 | - | \Lambda_2 |} \kappa^{\dist}
\la{formestimate7}
\eeq

(\ref{formestimate5}) and (\ref{formestimate7}) also hold 
for all $\Lambda_1' \subseteq \Lambda_1$,
$\Lambda_2' \subseteq \Lambda_2$ and (1.) follows.

\ba
\pi_{\Lambda_1}(f \nu ) 
& = & 
\pi_{\Lambda_1} (f \nu_{\Lambda_1 \cup \Lambda_2})\\
& = &
\pi_{\Lambda_1}(f \nu_{\Lambda_1} \nu_{\Lambda_2}
+ f(\nu_{\Lambda_1} \nu_{\Lambda_2} -
\nu_{\Lambda_1 \cup \Lambda_2}))\nn\\
& = & 
\nu (f) \nu_{\Lambda_1} + \pi_{\Lambda_1}
(f(\nu_{\Lambda_1} \nu_{\Lambda_2} -
\nu_{\Lambda_1 \cup \Lambda_2}))
\nn
\ea

and using $\|\pi_{\Lambda_1}\|_\infty =1$ we get

\beq
\|\pi_{\Lambda_1} (f(\nu_{\Lambda_1} \nu_{\Lambda_2} -
\nu_{\Lambda_1 \cup \Lambda_2}))
\|_{\Lambda_1}
\leq
\| f\|_{\Lambda_2}
\|\nu_{\Lambda_1} \nu_{\Lambda_2} -
\nu_{\Lambda_1 \cup \Lambda_2}\|_{\Lambda_1\cup\Lambda_2}
\eeq

and so by (1.)

\beq
\|\pi_{\Lambda_1} (f(\nu_{\Lambda_1} \nu_{\Lambda_2} -
\nu_{\Lambda_1 \cup \Lambda_2}))
\|_{\Lambda_1}
\leq
c_{16} \vth^{- |\Lambda_1|- |\Lambda_2|} \| f\|_{\Lambda_2} 
\kappa^{\dist}
\eeq

This holds for all $\Lambda_1' \subset \Lambda_1$, so
(2.) is proved.

We set $\phi = f \nu - \nu (f) \nu$. So
$\pi_{\Lambda_1}\circ\mathcal{L}^N_{F
\circ T^{\e}} (f\nu ) - \nu (f) \nu_{\Lambda_1}
= \pi_{\Lambda_1}\circ\mathcal{L}^N_{F
\circ T^{\e}}\phi$.
We estimate the 
$\|\cdot\|_{\Lambda_1 ,\tvth}$-norm of the
last term as in the proof of 
Proposition \ref{propetalaw}, but this time using the finer estimates
from (2.)

\ba
\|\phi_{\Lambda (\mathcal{C})}\|_{\Lambda (\mathcal{C})}
& \leq &
\vth^{-|\Lambda (\mathcal{C})|} c_{11} \vth^{-|\Lambda_2|}
\| f\|_{\Lambda_2}
\kappa^{\mbox{\tiny dist}(\Lambda (\mathcal{C}) ,\Lambda_2)}\\
& \leq &
c_{11} \vth^{-|\Lambda_2|} \| f\|_{\Lambda_2}
\vth^{-| \Lambda_r (\mathcal{C}) |
-\sum_{k=1}^\infty (3k)^d n_{\beta ,k}}
\kappa^{{\dist} - \sum_{k=1}^\infty k n_{\beta ,k}}
\nn
\ea

where as before 
$\Lambda (\mathcal{C}) \stackrel{\mathrm{def}}{=}
\tilde{\Lambda}_{\mathcal{C}} \cup \Lambda_r$.

So we get analogously to formulae 
(\ref{formestimate0}) and (\ref{formestimate2}):

\ba
\lefteqn{
\tvth^{|\Lambda_1 |}
\sum_{\mathcal{C}: \mbox{\tiny length}(\mathcal{C})=N}
\left\| \pi_{\Lambda_1} \circ\mathcal{L_C} \phi_{\Lambda_2}
\right\|_{\Lambda_1}}\\
& \leq &
\tvth^{|\Lambda_1 |}
\sum_{K=0}^{|\Lambda_1 |} {|\Lambda_1 | \choose K}
\sum_{n_\beta \atop K\leq |n_\beta|<\infty} 4^K
\prod_{k=1}^{\infty} \left(\exp (c_{21} k^d)\right)^{n_{\beta ,k}}
(c_1 \e)^{|n_\beta|}
\nn\\
&& \, \times
\prod_{k=1}^{\infty} \left(\exp (-c_2 k^d)\right)^{n_{\beta ,k}}
\sum_{L=|n_\beta |}^\infty {L \choose |n_\beta |}
c_r^{|n_\beta |} \eta^L
\prod_{k=1}^{\infty} \left(\exp (c_{22} k^d)\right)^{n_{\beta ,k}}
\nn\\
&& \,\times
\prod_{k=1}^{\infty} \left(\exp (c_{23} k^d)\right)^{n_{\beta ,k}}
\eta^{\max \{ 0,N-L\}}
\sum_{l=0}^{|\Lambda_1 | -K} {|\Lambda_1 | -K \choose l}
(c_r\eta^N)^l c_h^{|\Lambda_1 | -K-l} 
\nn\\ 
&&\,\times
c_{11} \vth^{-|\Lambda_2|} \| f\|_{\Lambda_2}
\vth^{-l-\sum_{k=1}^\infty (3k)^d n_{\beta ,k}(\mathcal{C})}
\kappa^{{\dist} - \sum_{k=1}^\infty k n_{\beta ,k}}
\nn\\
& \leq &
c_{11} \tvth^{|\Lambda_1 |}
\sum_{K=0}^{|\Lambda_1 |} {|\Lambda_1 | \choose K}
\sum_{n_\beta \atop K\leq |n_\beta|<\infty} 4^K
(c_1 \e c_r)^{|n_\beta|}
\nn\\
&& \,\times
\prod_{k=1}^{\infty} \left(\exp ((c_{21}-c_2+c_{22}+c_{23}-3^d\ln\vth 
-\ln\kappa )k^d)\right)^{n_{\beta ,k}}
\nn\\
&& \,\times
\sum_{L = n_\beta}^\infty \eta^{\max \{ L,N \}}
(\vth^{-1}c_r \eta^N +c_h)^{|\Lambda_1| -K}
\vth^{-|\Lambda_2|} \| f\|_{|\Lambda_2|} \kappa^{\dist}
\nn
\ea

This time we set
$\tilde{c_2}=
c_2 -c_{21}-c_{22}-c_{23}+3^d\ln\vth +\ln\kappa$ and with the same 
analysis as from 
(\ref{formestimate2}) to (\ref{formestimate9}) we get:

\beq
\leq 
c_{34}
(\e_2 + c_r \tilde{\eta}^N \frac{\tilde{\vth}}{\vth}
+c_h \tilde{\vth})^{|\Lambda_1 |}
\|\nu\|_\vth \vth^{-|\Lambda_2|} \| f\|_{\Lambda_2} \kappa^{\dist}
\tilde{\eta}^N .
\la{formestimatekappadist1}
\eeq

For sufficiently small $\e_2$ and $\tvth$ the term in brackets
is smaller than one. Note that there is no condition on $N$.
So we get the same estimates for all $n \geq 0$ and these also hold 
for $\Lambda \subset \Lambda_1$. So in analogy with
(\ref{formcauchycritpfo}) we get

\beq
\left\|\mathcal{L}^N_{F \circ T^\e} \phi
- \mathcal{L}^{N+1}_{F \circ T^\e}\phi \right\|_{
\Lambda_1}
\leq c_{35} \vth^{-|\Lambda_2|} \| f\|_{\Lambda_2} \kappa^{\dist}
\tilde{\eta}^N
\eeq

and as $\mu (\phi )=0$ we conclude (3.)
\qed


{\bf Proof of Theorem \ref{theoremdecay2}}
\ba
\lefteqn{
\left| \int_M \nu d\mu \, g f - 
\left( \int_M \nu d\mu \, g \right)
\left( \int_M \nu d\mu \, f \right) \right|}\\
& \leq &
\left|
\int_{(S^1)^{\Lambda_1\cup\Lambda_2}}
d \mu^{\Lambda_1\cup\Lambda_2} 
(\nu_{\Lambda_1\cup\Lambda_2} -
\nu_{\Lambda_1} \nu_{\Lambda_2}) g f \right|\nn\\
& \leq &
\|\nu_{\Lambda_1\cup\Lambda_2} - 
\nu_{\Lambda_1} \nu_{\Lambda_2}\|_{\Lambda_1\cup\Lambda_2}
\| g\|_\infty \| f\|_\infty\nn\\
& \leq &
c_{10} \vth^{-|\Lambda_1|-|\Lambda_2|}
\| g\|_\infty \| f\|_\infty 
\kappa^{\mbox{\tiny dist} (\Lambda_1 ,\Lambda_2 )}
\nn
\ea

so (1.) is proved.

\ba
\lefteqn{
\left| \int_M \nu d\mu \, g\circ\tau\circ S^n f - 
\left( \int_M \nu d\mu \, g\circ\tau \right)
\left( \int_M \nu d\mu \, f \right) \right|}\\
& = &
\left| \int_M d\mu \, g\circ\tau
\left(\pi_{\tau^{-1}(\Lambda_1 )} \circ \mathcal{L}^n_{F \circ T^\e}
(f\nu) - \nu (f) \nu_{\tau^{-1}(\Lambda_1 )}
\right)\right|\nn\\
& \leq &
c_{12} c_5^{|\Lambda_1|+|\Lambda_2|}\| f\|_{\Lambda_2} 
\| g\|_\infty
\kappa^{\mbox{\tiny dist}(\tau^{-1}(\Lambda_1), \Lambda_2 )} 
\tilde{\eta}^n
\nn
\ea

Here we have used (3.) of Theorem \ref{propositiondecay1}
and set $c_5 \stackrel{\mathrm{def}}{=} \tvth^{-1}$.
From

\beq
\mbox{dist}(\tau^{-1}(\Lambda_1), \Lambda_2 ) \geq m(\tau ) 
- \mbox{diam} (\Lambda_1 ,\Lambda_2 )
\eeq

follows

\beq
\kappa^{\mbox{\tiny dist}(\tau^{-1}(\Lambda_1), \Lambda_2 )}
\leq 
c(\Lambda_1 ,\Lambda_2 ,\kappa) \kappa^{m(\tau )}
\eeq
 
where $c(\Lambda_1 ,\Lambda_2 ,\kappa)$
depends only on $\Lambda_1$, $\Lambda_2$ and $\kappa$.
If $\tau$ and $S$ commute, (3.) follows from (2.).

We prove (4.) by approximating $g$ and $f$ by functions and for that we can
apply estimate (2.).
For a given $\gamma >0$ we choose
$\Lambda_1\in\mathcal{F}$ so large that 
$\|g - g_{\Lambda_1}\|_\infty \leq \gamma$.
Further there exists an
$\tilde{f}_{\Lambda_2}\in \mathcal{H}(\adz )$ with
$\|f - \tilde{f}_{\Lambda_2}\|_\infty\leq\gamma$
(sup-norm on $(S^1)^{\Z^d}$).
So

\ba
\lefteqn{
\left| \int_M \nu d\mu \, g\circ\tau\circ S^n f - 
\left( \int_M \nu d\mu \, g\circ\tau \right)
\left( \int_M \nu d\mu \, f \right)\right|}\\
& \leq &
\left|
\int_M \nu d\mu \, (g - g_{\Lambda_1}) \circ\tau\circ S^n f
\right|
+
\left|
\int_M \nu d\mu \, g_{\Lambda_1}\circ\tau\circ S^n 
\, (\tilde{f}_{\Lambda_2} - f)
\right| \nn\\
& & +
\left|
\int_M \nu d\mu \, g_{\Lambda_1}\circ\tau\circ S^n \tilde{f}_{\Lambda_2} - 
\left( \int_M \nu d\mu \, g_{\Lambda_1}\circ\tau \right)
\left( \int_M \nu d\mu \, \tilde{f}_{\Lambda_2} \right)
\right| \nn\\
& & +
\left|
\left( \int_M \nu d\mu \, g_{\Lambda_1}\circ\tau \right)
\left( \int_M \nu d\mu \, (f - \tilde{f}_{\Lambda_2}) \right)
\right| \nn\\
& & +
\left|
\left( \int_M \nu d\mu \, (g - g_{\Lambda_1})\circ\tau \right)
\left( \int_M \nu d\mu \, f \right)
\right| \nn\\
& \leq &
\|g - g_{\Lambda_1}\|_\infty \|f\|_\infty 
+ \|g_{\Lambda_1}\|_\infty \|f - \tilde{f}_{\Lambda_2}\|_\infty \nn\\
& & + 
c(\Lambda_1 ,\Lambda_2 ,\kappa)
c_5^{|\Lambda_1|+|\Lambda_2|} \| g_{\Lambda_1}\|_\infty
\| \tilde{f}_{\Lambda_2}\|_{\Lambda_2} 
\tilde{\eta}^{n(\sigma )} \kappa^{m(\sigma )} \nn\\
& & +\|g_{\Lambda_1}\|_\infty \|f - f_{\Lambda_2}\|_\infty
+\|g - g_{\Lambda_1}\|_\infty \|f_{\Lambda_2}\|_\infty \nn\\
& \leq &
\left(
2 \| f\|_\infty +2 \| g\|_\infty + 3 \gamma
\right)
\gamma \nn\\
& & 
+ c(\Lambda_1 ,\Lambda_2 ,\kappa)
c_5^{|\Lambda_1|+|\Lambda_2|}
(\| g \|_\infty + \gamma ) 
\| \tilde{f}_{\Lambda_2}\|_{\Lambda_2}
\tilde{\eta}^{n(\sigma )} \kappa^{m(\sigma )}
\nn
\ea

and this gets arbitrarily small as we first choose $\gamma$, then
$\Lambda_1$, $\Lambda_2$ and $f_{\Lambda_2}$ and finally
$\max \{ m(\sigma ), n(\sigma )\}$.

(5.) follows from (4.) and the commutation of the $\tau_{e_i}$ with $S$.
\qed


\section*{Acknowledgements}
The first author would like to thank T. Hefer for helpful discussions
and R. Reid for his comments.
He is supported by the EC via the TMR-Fellowship ERBFMBICT-961157.



\begin{thebibliography}{99}

\bibitem{baladi} V. Baladi, M. Degli Esposti, S. Isola, E. 
J\"{a}rvenp\"{a}\"{a}, A. Kupiainen:
The spectrum of weakly coupled map lattices, preprint

\bibitem{bricmontkupiainen1} J. Bricmont and A. Kupiainen:
Coupled Analytic Maps, \emph{Nonlinearity} 8, pp.379-396, 1995

\bibitem{bricmontkupiainen2} J. Bricmont and A. Kupiainen:
High Temperature Expansions and Dynamical Systems,
\emph{Comm. Math. Phys.} 178, pp.73-732

\bibitem{bricmontkupiainen3} J. Bricmont and A. Kupiainen:
Infinite dimensional SRB-measures,
preprint

\bibitem{bunimovich} L.A. Bunimovich:
Coupled map lattices: One step forward and two steps back,
\emph{Physica D 86}, 1995, pp. 248-255 

\bibitem{bunimovichsiani} L.A. Bunimovich, Y.G. Sinai:
Space-time chaos in coupled map lattices, \emph{Nonlinearity 1},
pp. 491-516, 1988

\bibitem{dunfordschwartz} N. Dunford and J.T. Schwartz: 
Linear Operattors Part I, \emph{Interscience}, New York, 1958

\bibitem{jiang1} M. Jiang:Equilibrium states for lattice models of 
hyperolic type
\emph{Nonlinearity 8}, no.5, 1994, pp.631-659

\bibitem{jiang2} M. Jiang: Ergodic Properties of Coupled Map Lattices of
Hyperbolic Type, 
\emph{Penns. State University Dissertation}, 1995

\bibitem{jiangmazel} M. Jiang and A. Mazel: uniqueness of Gibbs States
and exponential Decay of Correlation for some Lattice models,
\emph{Journal of Statistical Physics} 82, no. 3-4, 1995

\bibitem{jiangpesin} M. Jiang and Ya.B. Pesin:
Equilibrium Measures for Coupled Map Lattices: Existence, Uniqueness and
Finite-Dimensional Approximations, Preprint

\bibitem{kaneko} K. Kaneko (ed.): Theory and Applications of Coupled
map Lattices, \emph{J. Wiley}, 1993

\bibitem{kellerkuenzle} G. Keller and M. K\"{u}nzle:
Transfer Operators for Coupled Map Lattices, \emph{Ergodic Theory \&
Dynamical Systems} Vol. 12, pp.297-318, 1992

\bibitem{lasotamackey} A. Lasota and M.C. Mackey:
Chaos, Fractals and Noise, \emph{Springer}, 1994

\bibitem{maesmoffaert} C. Maes and A. Van Moffaert:
Stochastic Stability of Weakly Coupled Lattice Maps, preprint

\bibitem{pesinsinai} Ya.B. Pesin and Ya.G. Sinai:
Space-time chaos in chains of weakly interacting hyperbolic mappings,
\emph{Advances in Soviet Mathematics} Vol.3, 1991, pp.165-198

\bibitem{volevich1} D.L. Volevich: The Sinai-Bowen-Ruelle Measure
for a Multidimensional Lattice of interacting Hyperbolic Mappings,
\emph{Russ. Akad. Dokl. Math.} 47, 1993, pp.117-121

\bibitem{volevich2} D.L. Volevich: Construction of an analogue of 
Bowen-Ruell-Sinai measure for a multidimensional lattice of interacting 
hyperbolic mappings
\emph{Russ. Akad. Math. Sbornik} Vol.79, 1994, pp.347-363

\end{thebibliography}
\end {document}